\documentclass[journal]{IEEEtran}
\usepackage{amsmath,amsfonts}
\usepackage{algorithmic}
\usepackage{algorithm}
\usepackage{array}
\usepackage[caption=false,font=normalsize,labelfont=sf,textfont=sf]{subfig}
\usepackage{textcomp}
\usepackage{stfloats}
\usepackage{url}
\usepackage{verbatim}
\usepackage{graphicx}
\usepackage{cite}
\usepackage{booktabs}
\usepackage[most]{tcolorbox}
\usepackage{placeins}
\usepackage{setspace}

\begin{document}
\newcommand{\llamaeightname}[1]{\texttt{Meta-Llama-3-8B-Instruct}}
\newcommand{\llamaeight}[1]{Llama3 8B}
\newcommand{\llamaseventyname}[1]{\texttt{Meta-Llama-3-70B-Instruct}}
\newcommand{\llamaseventy}[1]{Llama3 70B}
\newcommand{\mixtralname}[1]{\texttt{Mixtral-8x7B-Instruct-v0.1}}
\newcommand{\mixtral}[1]{Mixtral 8x7B}
\newcommand{\mistralname}[1]{\texttt{Mistral-7B-Instruct-v0.3}}
\newcommand{\mistral}[1]{Mistral 7B}
\newcommand{\roberta}[1]{RoBERTa}
\title{Evaluating open-source Large Language Models for automated fact-checking}

\author{
    Nicolò Fontana\textsuperscript{\rm 1},
    Francesco Corso\textsuperscript{\rm 1,2},
    Enrico Zuccolotto\textsuperscript{\rm 1},
    Francesco Pierri\textsuperscript{\rm 1}\\
    \textsuperscript{\rm 1}Politecnico di Milano, Italy\\
    \textsuperscript{\rm 2}CENTAI, Italy\\
    \{nicolo.fontana, francesco.corso, francesco.pierri\}@polimi.it
}

\maketitle
\begin{abstract}
    The increasing prevalence of online misinformation has heightened the demand for automated fact-checking solutions. Large Language Models (LLMs) have emerged as potential tools for assisting in this task, but their effectiveness remains uncertain. This study evaluates the fact-checking capabilities of various open-source LLMs, focusing on their ability to assess claims with different levels of contextual information. We conduct three key experiments: (1) evaluating whether LLMs can identify the semantic relationship between a claim and a fact-checking article, (2) assessing models' accuracy in verifying claims when given a related fact-checking article, and (3) testing LLMs' fact-checking abilities when leveraging data from external knowledge sources such as Google and Wikipedia. Our results indicate that LLMs perform well in identifying claim-article connections and verifying fact-checked stories but struggle with confirming factual news, where they are outperformed by traditional fine-tuned models such as RoBERTa. Additionally, the introduction of external knowledge does not significantly enhance LLMs' performance, calling for more tailored approaches. Our findings highlight both the potential and limitations of LLMs in automated fact-checking, emphasizing the need for further refinements before they can reliably replace human fact-checkers.
\end{abstract}

\begin{IEEEkeywords}
fact-checking, large language models, prompting analysis
\section{Introduction}
\end{IEEEkeywords}

\label{sec:introduction}
In today’s digital era, the vast availability of online information has facilitated the rapid spread of both misinformation and disinformation. Online social platforms, in particular, enable false narratives to gain traction, due to their high-connectivity nature~\cite{DelVicarioMea16_SpreadingMisinformationOnline}. This challenge is amplified when influential individuals propagate misleading content, with research suggesting that such misinformation can significantly impact critical events, including election outcomes~\cite{FakeNewsInfluencingElections, BarDea24_SystematicDiscrepanciesDelivery}.

The burden of fact-checking and debunking false claims has been traditionally left to journalists.
However, the rise of the Internet, combined with growing distrust in traditional media~\cite{LackOfTrust}, has led to the emergence of independent fact-checking organizations~\cite{spivak2010fact}. 
These groups focus on verifying rumors, misconceptions, and fake news spread online.
One of the most notable fact-checking initiatives, Politifact\footnote{\url{https://www.politifact.com/}}, received the Pulitzer Prize for national reporting in 2009\footnote{\url{https://www.pulitzer.org/prize-winners-by-year/2009}}. 
Politifact introduced a rating system to evaluate claims, and many organizations have since adopted similar systems.
However, these subjective and often ambiguous ratings complicate comparisons across fact-checks~\cite{Lim2018}.

Fact-checking remains a labor-intensive process, requiring teams to spend days or even weeks verifying claims~\cite{WarrenGea25p_ShowMeTheWork}. Given the overwhelming flow of information online\footnote{\url{https://explodingtopics.com/blog/data-generated-per-day}}, traditional methods cannot keep pace, highlighting the need for more efficient approaches.

One promising avenue is the automation of fact-checking through artificial intelligence (AI) technologies~\cite{Guo2022}. Researchers have investigated AI-driven models, such as Convolutional Neural Networks (CNNs) and Graph Neural Networks (GNNs), to aid in these efforts. While these models have made significant progress, their effectiveness remains limited~\cite{HU2022133, GNN_all}. More recently, Large Language Models (LLMs) based on transformer architectures have demonstrated significant potential~\cite{LIN2022111, PLMs}. These models excel at generating natural language responses, answering complex questions, and producing high-quality content~\cite{hadi2023survey, FontanaNea24p_NicerThanHumans}, making them an accessible and versatile tool for a broad audience. Their advanced reasoning capabilities further position them as suitable candidates for fact-checking~\cite{PapageorgiouEea24_SurveyUseLarge}.

However, a major limitation of LLMs lies in the outdated nature of their training data~\cite{SanuEea24_LimitationsLargeLanguage}, which hampers their ability to address recent misinformation effectively. To address this challenge, researchers are exploring innovative approaches to integrate real-time, external knowledge into these models. Such advancements aim to enhance their ability to provide accurate and timely responses to rapidly evolving misinformation~\cite{Quelle_2024}.

Building on previous research, our study explores the use of diverse open LLMs to investigate whether smaller models can achieve a more favorable balance between cost-effectiveness and accuracy compared to larger models. Cost-effective methods are particularly critical for watchdog groups, non-profits, and smaller organizations that often operate with limited budgets and resources. These entities play a crucial role in combating misinformation but frequently lack the financial capacity to deploy and maintain large-scale AI systems. By identifying efficient yet accurate alternatives, our research aims to empower such organizations with accessible tools, enabling them to enhance their fact-checking capabilities without incurring prohibitive costs. Additionally, we employed as a baseline for comparison a Small Language Model (SLM) by fine-tuning \roberta{}. This reference allows us to understand the relative importance of the LLMs' performance.

We articulate our contributions into three key research questions:

\begin{enumerate}

\item \textbf{RQ1: Can LLMs identify the connection between a claim and an article?} This research question investigates whether LLMs can accurately determine if a given claim and its paired article are contextually related, addressing the same topic. The focus is on evaluating the models' capability to assess the relevance and alignment between the claim and the content of the article.

\item \textbf{RQ2: Can LLMs judge a claim based on the related fact-checking article?} This question explores whether LLMs can effectively analyze a related fact-checking article and provide a trustworthy evaluation of the claim based on the information contained within the document.

\item \textbf{RQ3: Are LLMs able to retrieve contextual information and fact-check claims?} This question assesses the models’ ability to verify the truthfulness of a claim when they are provided with a related article, as opposed to relying solely on their pre-trained internal knowledge. The objective is to determine whether the inclusion of new, external information enhances their accuracy in fact-checking tasks.

\end{enumerate}

To address the first two questions (\textbf{RQ1, RQ2}), we conduct experiments using 24 different prompts to guide the models’ performance, emphasizing the role of effective prompting.
For the third question (\textbf{RQ3}), we use neutral and straightforward prompts to evaluate various sources of external information--such as Google or Wikipedia (or their absence)--and the representation format (snippet, summary, or full article) that yields the best results.
In this third scenario, the model acts autonomously, conducting internet searches to gather information and refine its verdicts.
For Google searches, we limit the results to a time frame before the claim to avoid bias and better simulate real-world fact-checking.

Our analysis utilizes the Fact-Checking Insights dataset\footnote{\url{https://www.factcheckinsights.org/download}}, which is a comprehensive resource containing structured data from tens of thousands of claims made by political figures and social media posts, scrutinized and rated by independent fact-checking organizations such as AFP~\footnote{\url{https://factcheck.afp.com}}, Politifact~\footnote{\url{https://www.politifact.com/}}, and Snopes~\footnote{\url{https://www.snopes.com/}}.

\section{Related Work}
\label{sec:rel_work}

The study of fake news generation and dissemination has gained increasing attention, particularly with the rise of online platforms that facilitate rapid information diffusion~\cite{PierriFea19_FalseNewsOnSocialMedia}. Research in this domain has primarily focused on two key aspects: the identification of fake news and the detection of its spreaders~\cite{PapageorgiouEea24_SurveyUseLarge}.

Over time, human fact-checking has been increasingly supported by machine learning methods. Early approaches leveraged classical machine learning techniques for keyword-based text analysis using traditional neural networks~\cite{NasirJea21_FakeNewsDetection} and applied network analysis to assess the trustworthiness of sources propagating misinformation~\cite{MewadaAea24_CIPF}. More recently, the emergence of Large Language Models has significantly advanced fake news detection, enabling more sophisticated analyses. These include evaluating the performance of fake news detectors against synthetically generated misinformation rather than human-created content~\cite{SuJea24_AdaptingFakeNews} and conducting sentiment-based assessments through emotional analysis of news texts~\cite{ZhangXea21_MiningDualEmotion}.

Some approaches aim to enhance human fact-checking by integrating feedback from LLMs. For instance, LLMs have been used to improve document retrieval, aiding in the verification of statements~\cite{ZhangXea24_ReinforcementRetrievalLeveraging}, or to decompose claims hierarchically into sub-statements for systematic verification~\cite{ZhangXea23_LLMbasedFactVerification}. Conversely, other methodologies seek to achieve fully autonomous fact-checking, relying on LLMs to assess the veracity of claims using a zero-shot approach, without requiring additional training or human intervention.

A major challenge in automating fact-checking with LLMs lies in their tendency to produce biased answers~\cite{NogaraGea24p_ToxicBias, LiuGea25p_ComparingDiversityNegativityAndStereotypesInChinese-languageAITechnologies} and to hallucinate facts~\cite{AugensteinIea24_FactualityChallengesEra} and their reliance on static training data, which may often be outdated~\cite{SanuEea24_LimitationsLargeLanguage}. To address this limitation, researchers have explored methods to integrate external knowledge into LLMs using frameworks such as ReAct, which combines reasoning and action within LLMs~\cite{YaoSea23_ReAct}. This has led to several innovative fact-checking approaches. For example, FActScore assigns factuality scores to responses based on information retrieved from Wikipedia~\cite{MinSea23_FActScore}, while FACTOOL employs various tools for evidence collection and reasoning to analyze claims and assign factuality labels based on supporting evidence~\cite{ChernIea23p_FacTool}. Additionally, Toolformer enables LLMs to autonomously integrate external tools, enhancing performance across various tasks while preserving core language modeling capabilities~\cite{SchickTea23_Toolformer}. Despite these advancements, research indicates that while LLMs generally perform well, they struggle more with verifying factual statements than identifying false ones~\cite{Quelle_2024}.

To further improve accuracy, researchers have explored collaborative approaches where multiple models interact and reason together to reach a verdict. FactCheck-GPT, for instance, addresses factual inaccuracies by enabling multiple LLMs to debate and converge on a consensus through iterative discussions, supplemented by external searches such as Google~\cite{WangYea24_Factcheck-Bench, DuYea24_ImprovingFactualityReasoning, kim2024llmsproducefaithfulexplanations}.

Recently, the focus has expanded beyond mere verification to include misinformation correction. Systems like MUSE represent a significant advancement in this area, demonstrating the evolving capability of LLMs to both detect and correct misinformation in real-time environments~\cite{zhou2024correctingmisinformationsocialmedia}. Similarly, Verify-and-Edit enhances LLM reasoning by incorporating Chain-of-Thought (CoT) reasoning and external knowledge sources such as DrQA, Wikipedia, and Google Search to refine responses and correct factual inaccuracies~\cite{ZhaoRea23_Verify-and-Edit}. The Chain-of-Verification technique further improves factual accuracy by leveraging parametric knowledge to revise LLM-generated responses~\cite{DhuliawalaSea24_ChainofVerificationReducesHallucination}. Another approach, SELF-CHECKER, evaluates factuality by integrating real-time web searches (e.g., Bing) to assign factuality labels, reinforcing the role of external knowledge in automated fact-checking~\cite{LiMea24_Self-Checker}.

Contrary to prior research suggesting LLM superiority, some studies indicate that task-specific Small Language Models (SLMs), such as fine-tuned BERT, may outperform LLMs in certain fact-checking tasks. One study proposes a hybrid approach where LLMs generate rationales while SLMs handle classification, leveraging the strengths of both model types~\cite{HuBea24_BadActorGoodAdvisor}. This finding aligns with our analysis, which shows that LLMs fail to achieve significant performance improvements in zero-shot fact-checking on isolated statements, even when provided with temporally contextualized information. Similarly,~\cite{MaXea24_FakeNewsDetection} demonstrates that LLMs can be more effective as supplementary tools to traditional detection methods. Specifically, the study analyzes feature propagation in networks of entities and concepts extracted from articles using LLM assistance.

Another relevant research direction investigates the stylistic metrics underlying written content. One study explores whether LLMs exhibit human-like planning and creativity in news article generation~\cite{SpangherAea24_DoLLMsPlanLikeHumanWriters?}. Additionally, building on the \emph{Undeutsch} psychological theory that memories of real events differ from those of imagined ones,~\cite{WuJea24_FakeNewsInSheepsClothing} suggests that fake and real news may have distinct stylistic patterns. This raises the possibility that LLM-based fact-checking methods may rely more on stylistic cues than actual content analysis, particularly in low-context scenarios—an issue reflected in our final task.

Unlike other studies, we restrict the retrieval date settings in Google to obtain realistic estimates of how the LLM would perform on new data.
We also use a new dataset called Fact-Checking Insights, which compiles information from various sources to reduce bias. 
Instead of feeding articles to the LLM, we provide contextual data, such as the author and the date the claim was made.
Additionally, we conduct a temporal analysis of claim accuracy to determine whether the training data's cutoff date affects performance.
Furthermore, we focus on more economical, open-source models, limiting our analysis to those with at most 70 billion parameters to better understand their strengths and weaknesses.

\section{Experimental Design}
We focus exclusively on English-language claims, as most models are primarily trained in English and are expected to perform best in this language~\cite{lai-etal-2023-chatgpt}.
Our analysis utilizes the Fact-Check Insights dataset, a comprehensive resource available to researchers, journalists,  and other stakeholders engaged in countering political misinformation and falsehoods online.
This dataset comprises structured data from tens of thousands of claims made by political figures and social media posts, scrutinized and rated by independent fact-checking organizations such as AFP, Politifact, and Snopes.

\subsection{Data Preprocessing}

Starting from a collection of over 200K observations downloaded from Fact-Check Insights in June 2024, we removed duplicate entries, rows with incomplete or incorrect attributes, and data that were not in the English language. Next, we scraped the original fact-checking article associated with each claim in the dataset, removing those that could not be collected successfully.
After this process, we obtained a final dataset of 60.000 claims along with their corresponding fact-checking articles.

Following the literature~\cite{Quelle_2024}, we grouped the claims under two macro-categories: \textbf{True} and \textbf{False}. 
The former includes entries deemed accurate or of higher quality, while the latter contains content lacking a factual basis, such as fake quotes, conspiracy theories, misleading edited media, or ironic and exaggerated criticism.
In addition, we labeled all entries labeled as a mixture of true and false content as \textbf{False}. For example, labels like \emph{Geppetto mark}, \emph{trustworthy}, \emph{mostly true}, and \emph{correct attribution} were categorized as \textbf{True}, while labels such as \emph{false}, \emph{mostly false}, \emph{Quattro Pinocchio}, and \emph{unproven} or \emph{legend} were classified as \textbf{False}.
The resulting classes are highly unbalanced, with 90\% of the claims labeled as \textbf{False}.
This is reasonable since fact-checking activities usually focus on false claims~\cite{10.1145/3411763.3451760}. 

Next, we sampled 50 claims (25 \textbf{True} and 25 \textbf{False}) for each year from 2013 to 2023. We instead sampled 500 claims published in 2024, 30 of which only were \textbf{True} as this was the maximum number available in that year at the time of collection. 
We include observations from 2024 to better test the capabilities of LLMs at judging claims that are not available in their training data, i.e., because they were published after the training cutoff date of the model.

\subsection{Approach}
To evaluate LLMs as effective tools for fact-checking, we designed three distinct tasks aimed at thoroughly assessing their capabilities in this field and addressing the key questions outlined in Section~\ref{sec:introduction}.

\begin{enumerate}
    \item \textbf{Understanding the article-statement connection}: This task evaluates how accurately an LLM can answer when provided with a pair article-claim that are either related (i.e., the article is fact-checking the claim) or not (i.e., a random fact-checking article is picked).
    
    \item \textbf{Providing an accurate verdict based on a fact-checking article}: In this task, the model is required to answer what is the verdict on a given claim based on the associated fact-checking article.
    
    \item \textbf{Fact-Checking the claim}: In this task, the LLM evaluates the truthfulness of a given statement, with or without additional contextual information, using a neutral prompt (i.e. a prompt that does not steer the model towards a certain decision).
\end{enumerate}

Each task evaluates distinct aspects of the models' performance in fact-checking scenarios. We anticipate that the first two tasks will be relatively easier for the models, as they are required to answer queries based on manually provided contextual information.
These tasks test the models' capabilities using a range of prompts, as described below.
The third task instead simulates a real-world scenario where the model must retrieve relevant information to verify a given claim. To achieve this, we incorporate the ReAct framework~\cite{YaoSea23_ReAct}, enabling models to access the internet for gathering information useful in fact-checking.

In all cases, we provide the models with two basic metadata information related to the claim: the publication date and, when available ($\sim$70\% of the cases), the claim's author.
Including the date allows the model to contextualize the claim, as the validity of a statement may change over time with the emergence of new information. Knowing the claim's author can offer insights into its credibility, as claims from unreliable sources are more likely to be false~\cite{henkel2011reading}. Also, fact-checking agencies such as AFP and PolitiFact highlight the importance of both the claimant's identity and the claim's timing in assessing its veracity.

\subsection{Prompt Engineering}\label{subsec:prompt_engineering}
In this section, we provide a comprehensive overview of the 24 prompts evaluated in our experiments, which we categorize into three approaches: Zero-Shot, Few-Shot, and Chain-of-Thought. Zero-Shot prompts require the model to generate responses without any prior examples, relying solely on its pre-trained knowledge~\cite{WangWea19_ASurveyOfZero-ShotLearning}. Few-Shot~\cite{BrownTea20_LLMsAreFewShotLearners} prompts provide a small number of examples to guide the model toward the desired output style and reasoning process. Chain-of-Thought~\cite{wei2022chain} prompts explicitly break down the reasoning process into intermediate steps, encouraging the model to generate more structured and explainable responses. We construct these prompts by combining different modules, as detailed next.

\subsubsection{Basic prompt modules}
Here we detail a few basic components that are included in all prompts, in the order they appear as shown in Figure~\ref{fig:prompt_structure} and provided in the Appendix.

The \textbf{Role} prompt module primes the LLM to act as a fact-checker, aligning its reasoning with fact-checking principles. This framing encourages a more critical approach to evaluating claims and, as noted by \cite{KongAea24_BetterZeroShotReasoning}, can improve performance.

The \textbf{Task} prompt module requires the LLM to generate an explanation prior to providing an answer. This methodology is supported by literature indicating that additional tokens enhance the reasoning process of models~\cite{PelrineKea23_TowardsReliableMisinformationMitigation}. After generating the explanation, the model assigns a score (0-100) rather than a definitive verdict.
This scoring mechanism allows for flexible adjustment of the model's skepticism, as the acceptance threshold for responses can be modified.
We also decided to implement a scoring mechanism instead of relying on labels because, during the setup of our experiments, we empirically found that instructing the LLM to return a factual label often led to inconsistent outputs.
This approach is particularly relevant in our context, where understanding and verifying the reasoning behind a claim is crucial for ensuring the accuracy and trustworthiness of the verification process.

We employ \textbf{JSON} prompt module that provides a structured format for extracting expected responses, facilitating consistent and reliable evaluation across the various tasks. 
Adopting a similar approach to other work~\cite{json,PelrineKea23_TowardsReliableMisinformationMitigation}, we developed a prompt that enables the LLMs to produce structured JSON outputs.
Although these methods may occasionally result in null outputs (approximately 1\% of the results) they are highly advantageous, as they guarantee the presence of a result at the specified position.

The \textbf{Final} prompt module serves as a brief reminder to the model regarding how it should structure its responses, including the required format and the necessity of always providing a score.
This prompt is significant given that the articles being analyzed can be lengthy.
Smaller models, in particular, struggle to adhere to the specified format when processing long texts.
By including this reminder at the end, we significantly improve the model's precision in generating responses. Hence, this structured approach ensures that the model remains focused and compliant with the expected output format.

\begin{figure*}[!ht]
    \centering
    \includegraphics[width=\linewidth]{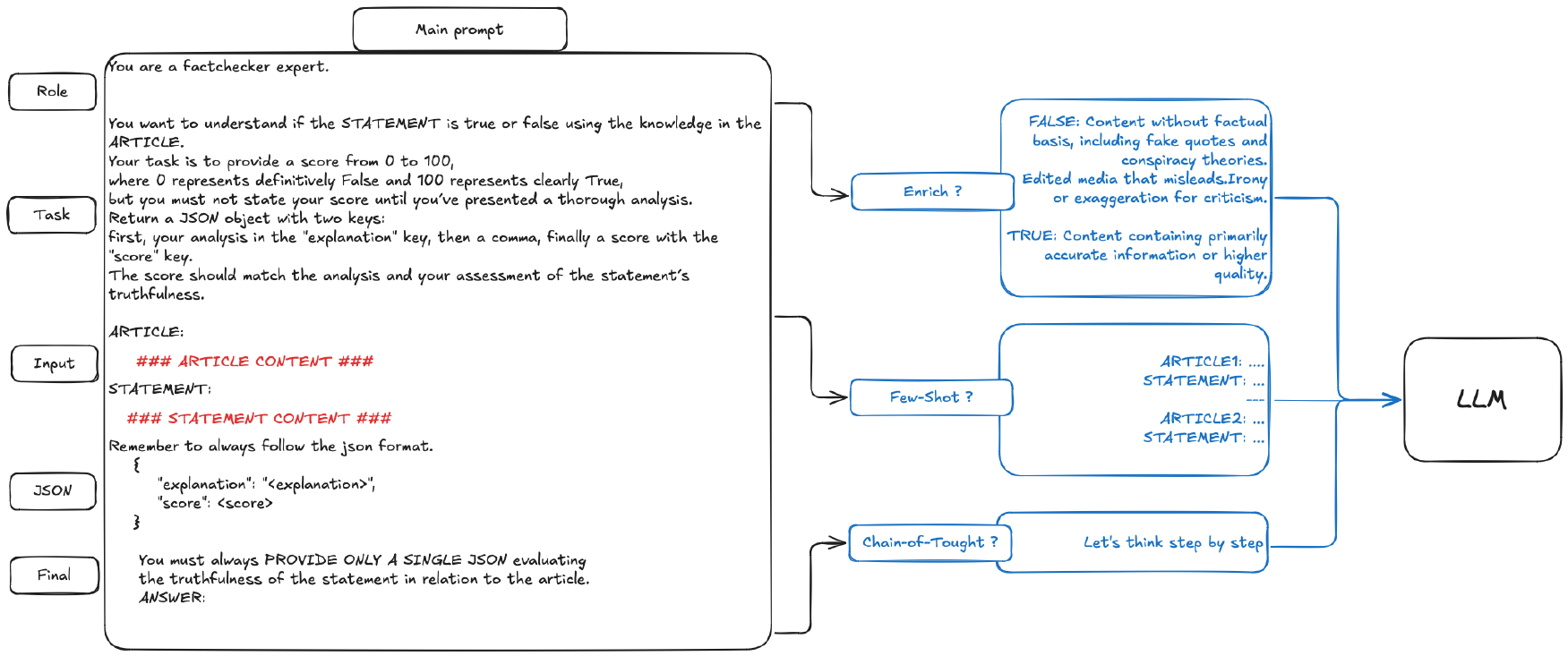}
    \caption{
    Example of prompt structure with optional configurations. The \emph{black} text represents the main prompt, shared across all tasks and configurations. \emph{Red} placeholders indicate where the actual article and statement are inserted for each task. \emph{Blue} components correspond to optional prompt modules, which are integrated into the main prompt (at the indicated position) based on the specific combination being tested. For instance, a prompt incorporating both \textbf{Enrich} and \textbf{Chain-of-Thought} will consist of the \emph{black} main prompt, with the \emph{blue} \textbf{Enrich} component placed between the \textbf{Role} and \textbf{Task} sub-prompts, and the \emph{blue} \textbf{Chain-of-Thought} component appended at the end.}
    \label{fig:prompt_structure}
\end{figure*}

\subsubsection{Prompt modules for Tasks 1 and 2}
We consider three different approaches for the first two tasks:

\begin{itemize}
  \item \textbf{Zero Shot (ZS)}: This prompt module requires the model to generate responses based solely on its preexisting knowledge and understanding.
  It does not require examples and it relies on indirect information for prediction. However, it heavily depends on the quality of semantic information and may struggle with highly diverse categories~\cite{schulhoff2024prompt}.
  \item \textbf{Few Shot (FS)}: Unlike the ZS approach, the FS prompt module provides the model with several examples before responding. In this process, we provide one example per class to avoid the limitations of one-shot prompting, which relies on only one example. However, there is a risk of overfitting the small dataset, and the model's performance can be susceptible to the quality and diversity of the examples provided. 
  To minimize these problems, the instances are randomly selected in each iteration.
  Also, to avoid confusion, especially for smaller models, we encapsulated each sample within square brackets to emphasize that these represent distinct entities, leading to more precise and clearly defined responses from the model.
  The example articles were limited to remain inside the context windows.
  This method assesses the model's ability to generalize from a limited set of examples~\cite{schulhoff2024prompt}.
  \item \textbf{Chain of Thoughts (CoT)}: Recent research has shown that adding a chain-of-thought paradigm can significantly boost the performance of language models across various tasks~\cite{zhou2023leasttomostpromptingenablescomplex, SuzgunMea23_ChallengingBIGBenchTasks}. In this approach, we include the phrase "Let's think step-by-step" after the examples, recognizing that language models can act as zero-shot reasoners~\cite{kojima2022large}. This technique aims to enhance reasoning skills and achieve better results.
\end{itemize}

Furthermore, we added the following approach to obtain more sophisticated prompts:

\begin{itemize}
    \item \textbf{Enriched criteria}: This approach enriches the prompt by providing a more explicit rationale behind labels. For the \textit{False} category, we specify that it includes misinformation types such as fake quotes and conspiracy theories. Conversely, the \textit{True} category encompasses content that demonstrates factual accuracy or contains higher-quality information. As suggested by previous studies, this enhancement is expected to lead to higher accuracy in ambiguous cases. In our context, it should assist the model in accurately labeling mixed verdicts, such as \textit{mixture}~\cite{zhou2024correctingmisinformationsocialmedia}.
    \item \textbf{Self-Reflection}: Once we receive the model's answer, we feed the LLM with the previous prompt and its generated response and ask it to reflect on its answer. This allows the model to reason and think about its answer before providing a new response. By encouraging this reflective process, we aim to reduce instances of hallucination, where the model might generate inaccurate or fabricated information. This approach fosters critical thinking and should enhance the reliability of the model's outputs~\cite{ji-etal-2023-towards}.
 \item \textbf{Summary}: We also investigated whether providing a summary focused on the claim rather than an entire article could enhance the effectiveness of our approach. This strategy distills relevant information, optimizing the context window's use. This prompt depends hardly on the large language model's ability to generate a coherent and accurate summary.
\end{itemize}
The last two methods are more costly because they involve calling the LLM twice and require processing longer texts compared to the first method. In contrast, the first method involves only a minimal increase in tokens, making it far more cost-efficient.

\subsubsection{ReAct for Task 3}
Our fact-checking methodology incorporates the ReAct framework~\cite{YaoSea23_ReAct} to enhance precision and efficiency in information retrieval. 
This method can be broken into two pieces: Reasoning (Re) and Act (Action).
First, the model uses its internal knowledge to reflect and think about the user's input, and then it identifies the steps required to solve the problem.
After Reasoning, the model starts to Act following the steps identified before; in our case, it acts using an external tool to retrieve information from a source. 
This framework allows the model to perform this series of Reasoning an Action multiple times, but we limited our experiment to only one iteration.
The model could also conclude that it does not need to search online to produce an optimal result and, after just one reasoning step, return the answer; otherwise, it would perform one search before answering.
In this framework, the LLM operates using a structured conversational ReAct JSON prompt, ensuring that each response adheres to a predefined format. The JSON responses can take one of two forms: "Final answer," which includes the final score and explanation, or "Action," which enables the model to perform an online search.
When performing a search, the agent autonomously generates a query and calls the search tool, providing the query as metadata.
After the tool retrieves relevant information, the LLM is called again to produce a final answer, incorporating the newly obtained knowledge.

In particular, we utilize two information sources: Wikipedia and Google. 
For Wikipedia, we leverage its API by developing a tool that, given a query generated by the LLM, returns the top 3 matches. Each match includes the title, link, and a snippet. These results are used to retrieve the snippets, while for complete content retrieval, we scrape the corresponding URLs, limiting the data to a certain length to ensure it fits within the context window.
Finally, to generate summaries, we feed the scraped pages back into the model and request a summary for each page, resulting in three separate model calls for summarization.

Meanwhile, for Google, we simulate an actual user experience using the Selenium automation tool. We apply a stricter criterion, filtering results based on a time range of up to one week before the claim's assertion date.
Using Selenium, we systematically scan the first 20 search results and extract the top three relevant links.
As with Wikipedia, we use the results to gather snippets.
For full content, we scrape the URLs, and in the case of summaries, we prompt the model to generate summaries from the scraped pages.

Following this process, we conduct experiments across three different settings:

\begin{enumerate} \item \textbf{Snippets}: In this setting, we provide only short snippets of information. This approach allows for a quick, surface-level analysis and is highly cost-effective.

\item \textbf{Full Article}: Here, we present the complete articles retrieved from the search results. This setting enables a more in-depth content analysis but may risk overloading the model's context window.

\item \textbf{Summary}: In this approach, we supply LLM-generated summaries based on the articles collected in the previous setting. These summaries should distill the essential information, facilitating efficient analysis and comparison while avoiding the context window limitation. However, this method is more expensive as it requires invoking the LLM twice. \end{enumerate}
As we can see in this task, the number of operations, the complexity, the strict formatting requirements, and the use of external tools make it more challenging for the model to follow the format consistently. As a result, we observed that approximately 2\% of the outputs were invalid.

All the prompts used in this study are detailed in the Appendix.

\subsection{Experimental settings}
In our study, we employed four models: two small and two large ones. Specifically, we utilized the Mistral family models, including \mistralname{} and \mixtralname{}, alongside LLaMA models, which included \llamaeightname{} and \llamaseventyname{}.

For interaction with these models, we resort to the \textit{Hugging Face API}~\footnote{\url{https://huggingface.co/blog/inference-pro#supported-models}}.
We established fixed parameters to ensure reproducibility across all tasks and tested models.
We set the temperature to a low value of 0.1 to enhance the consistency of our experiments, while the number of new tokens was fixed at 256.

To perform the experiments for Task 1 and Task 2, we structured a framework with placeholders, as presented in subsection~\ref{subsec:prompt_engineering}.
We experimented with all possible combinations of different prompt techniques for each dataset entry.

For the Summary enhancement, we provided a summary instead of feeding the full article to the model.
As mentioned before, we asked the model to generate a separate response, following the steps outlined earlier.
These procedures were repeated for each model across the first two tasks.

\subsubsection{Task 1: Understanding the article-statement connection}
Our experiments involved feeding our dataset into the models. Each entry was resource-intensive due to its 24 settings.
To ensure a balanced dataset, we divided the entries into two groups: one with the correct claim and the other with a random, unrelated claim.
This allowed us to create a dataset with an equal number of \textbf{explained} and \textbf{unexplained} entries.

\subsubsection{Task 2: Providing an accurate verdict based on the article's knowledge}
Similarly to the previous task, this experiment involved a dataset of 1,000 samples.
We provided the LLM with the article verifying the claim and asked for their verdict, which was explicitly contained within the article.
Our primary objectives were to verify the accuracy of our label mapping and to assess whether the LLMs could effectively extract critical information from the text.
This task was instrumental in refining our label mapping.

\subsubsection{Task 3: Fact-Checking the claim}
In this task, we utilized the complete dataset.
Unlike the previous tasks, we provided the model with claims along with their corresponding metadata and allowed it to explore freely in 3 settings: Wikipedia, Google, or no context.

As described in previous sections, we employed a ReAct agent that required the LLM to perform queries and interact with external resources, specifically the Wikipedia API and the Selenium tool. 

We developed an agent using the \textit{Langchain} \footnote{\url{https://www.langchain.com}} library to implement this functionality, creating two specific tools: one for Wikipedia and another for Google.

By cleverly manipulating the URL, we filtered out results published. This approach aimed to avoid articles that clearly fact-check the claim.

We deployed our models as autonomous agents by integrating the capabilities of LangChain with its HuggingFace integration.
To ensure objectivity in the responses, we employed a zero-shot prompt while assigning the model the role of a fact-checker.
We evaluated three scenarios by providing the model with different input formats for each source of information: the snippet, the full article, or a summary generated by the LLM.

\subsection{Extraction of results}
While extracting results from the model, we encountered situations where the model returned multiple answers for a single query.
In such cases, we selected the last complete answer, which was often the most comprehensive.
To ensure the extracted data complied with the JSON format, we addressed formatting issues, specifically converting single quotes (') to double quotes ("). This adjustment was made using a regular expression.

Once the formatting was corrected, we utilized the JSON structure to extract relevant scores and explanations.
This data was then compiled into a resulting dataset and subsequently fed into another program for score computation.

We recorded \textit{None} entries in the dataset when the extraction failed or yielded no valid response.

We decided to fix the threshold for a positive label with a score of 50.
Labels with a value of 50 or lower were categorized as \textit{False}, while those greater than 50 were categorized as \textit{True}.
Any values outside the range of 0 to 100 were classified as \textit{None}.

\subsection{Baseline}
To compare our models, we fine-tuned a \roberta{} model to perform binary classification on article-claim pairs (Task 1 and Task 2) or claim-only embeddings (Task 3) using our dataset. This dataset consists of a large number of claims, each labeled as \textbf{True} or \textbf{False}, and originally paired with fact-checking articles. However, given that \roberta{} can only process sequences up to 512 tokens, we decided to reduce the articles' length, which can exceed 20,000 characters.

For preprocessing, we tokenized the claims using the \roberta{} tokenizer with a maximum sequence length of 512 tokens. We trained the model on this dataset using a cross-entropy loss function and an AdamW optimizer, fine-tuning for three epochs. The dataset was split into training and validation sets, and model performance was evaluated using accuracy and classification metrics.

Our results indicate that fine-tuning \roberta{} shows great performance in distinguishing between true and false claims.

These results are included as baselines in the main graphs.

\subsection{Evaluation metrics}

To evaluate our model's performance comprehensively, we employ five distinct metrics:  Precision, Recall, F1 score, and ROC AUC Score.
\begin{itemize}
    \item \textbf{Recall (Sensitivity/True Positive Rate)}: Measures the proportion of actual positives that were correctly identified. 
    \[
    \text{Recall} = \frac{\text{TP}}{\text{TP} + \text{FN}}
    \]

    \item \textbf{Precision (Positive Predictive Value)}: Measures the proportion of predicted positives that are truly positive. 
    \[
    \text{Precision} = \frac{\text{TP}}{\text{TP} + \text{FP}}
    \]

    \item \textbf{Accuracy}: The overall proportion of correct predictions, both true positives and true negatives. 
    \[
    \text{Accuracy} = \frac{\text{TP} + \text{TN}}{\text{Total Population}}
    \]

    \item \textbf{F1 Score}: The harmonic mean of precision and recall. It balances the two, which is especially useful when there is an uneven class distribution like in our case. 
    \[
    \text{F1 Score} = 2 \times \frac{\text{Precision} \times \text{Recall}}{\text{Precision} + \text{Recall}}
    \]

    \item \textbf{ROC AUC Score (Area Under the Curve)}: A numerical value representing the area under the ROC curve.
    It summarizes the model’s ability to distinguish between classes.
    A score of 1.0 indicates perfect classification, while 0.5 indicates random guessing.
\end{itemize}

These metrics collectively offer a detailed understanding of the model's efficacy. In addition, we compute recall, precision, and f1 score separately for each class, thereby distinguishing the model's handling of true positives/negatives from false positives/negatives. This differentiation is crucial for identifying any potential difficulties the model may encounter when processing different cases.

In fact-checking, the disaggregation of evaluation metrics cases serves a particularly critical function. This distinction is important because the implications of misclassification can vary considerably depending on whether a true statement is incorrectly labeled as false or a false statement is incorrectly labeled as true. It also helps identify potential biases within the model, such as a predisposition towards skepticism or credulity. Furthermore, this strategy aligns closely with the practical priorities of fact-checking, where the relative importance of avoiding false positives versus false negatives can vary depending on the context.

\section{Results}
We present results for \llamaeight{}, \llamaseventy{}, and \mixtral{}, while the results for \mistral{} have been relegated to the Appendix due to the exceptionally high number of faulty responses generated by this model. Interestingly, this issue was further exacerbated when a token limit was applied to the answer prompt, while, this same modification reduced the occurrence of faulty responses in the other models.

\subsection*{Task 1: Understanding the article-statement connection}

\begin{figure}[!ht]
    \centering
    \includegraphics[width=\linewidth]{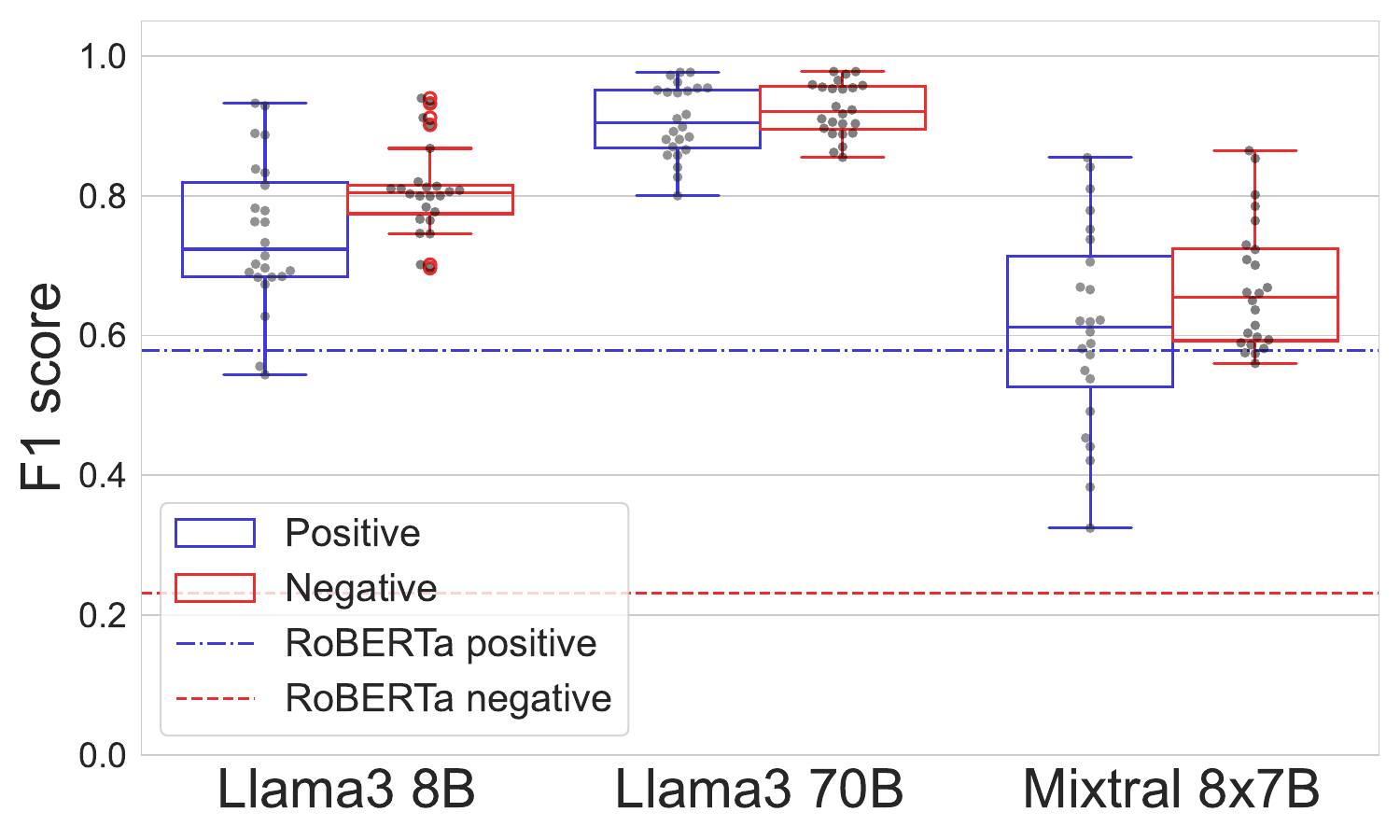}
    \caption{Task 1: Models' F1 scores computed for both classes. Fine-tuned \roberta{} is used as a reference baseline.}
    \label{fig:t1_f1_box}
\end{figure}

\begin{figure}[!ht]
    \centering
    \includegraphics[width=\linewidth]{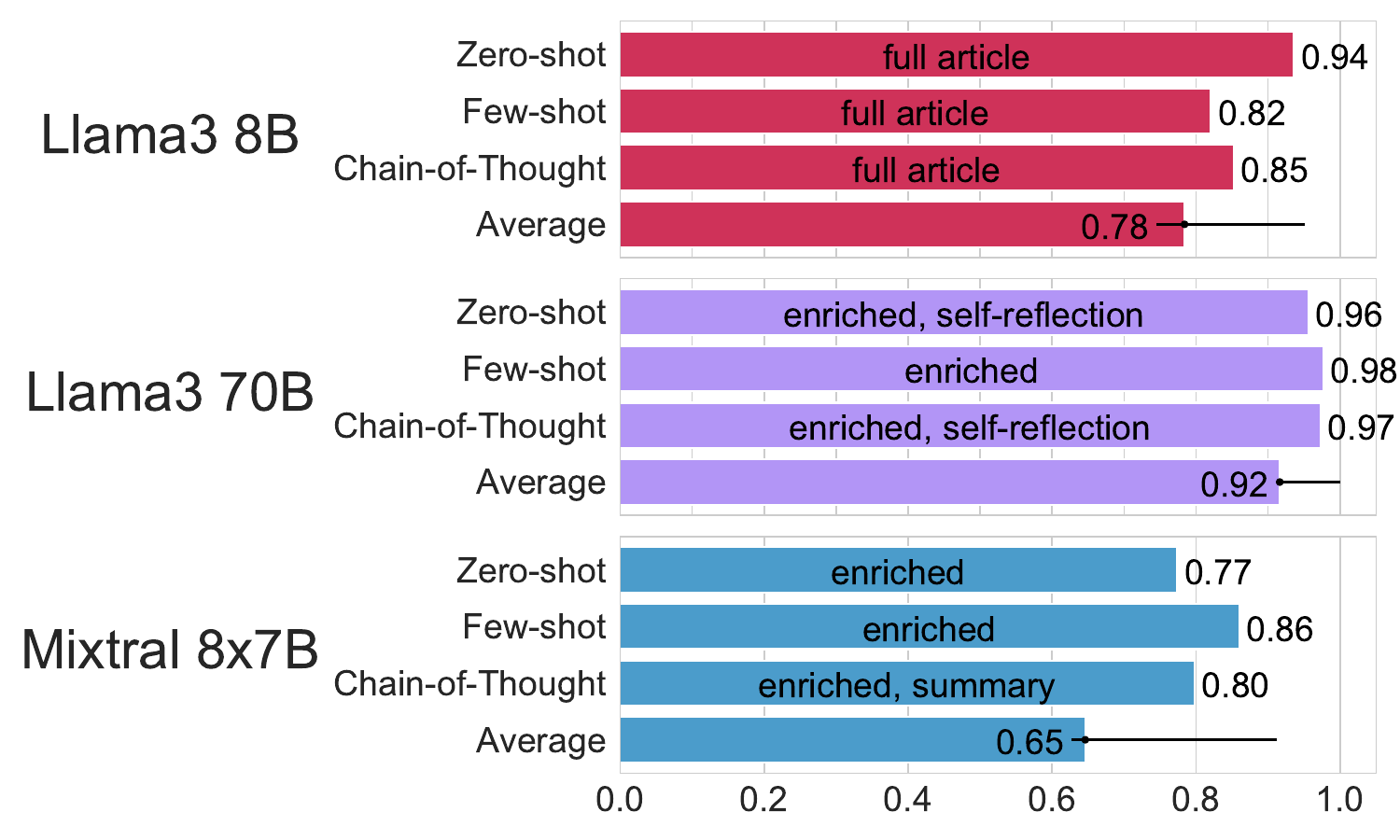}
    \caption{Task 1: For each model, for each prompt category (Zero-Shot, Few-Shot, and Chain-of-Tought) the best prompt's F1 score is reported. The average F1 score for each model is also reported with 0.95 confidence intervals.}
    \label{fig:t1_best_prompts}
\end{figure}

In Fig.~\ref{fig:t1_f1_box}, we show that all models outperform the fine-tuned \roberta{}, which exhibits particularly low F1 scores, especially for the negative class. The only exceptions are certain prompts for both \mixtral{} and \llamaeight{}. Among the three models, \llamaseventy{} consistently delivers the strongest performance, reaching an F1 score of approximately 0.9, depending on the prompt configuration. In contrast, \llamaeight{} and \mixtral{} produce lower scores, with \mixtral{} emerging as the weakest model, averaging 0.65 (Fig.~\ref{fig:t1_best_prompts}) and occasionally dropping below the random threshold of 0.5.

As shown in Fig.~\ref{fig:t1_best_prompts}, and consistent with findings in the literature~\cite{corso2024conspiracy}, no single prompt format stands out as the best across all models. For \llamaeight{}, providing the full article without relying on more complex prompting strategies yields the highest F1 scores across all prompt configurations. Meanwhile, \llamaseventy{} and \mixtral{} achieve their best results when enriched prompts—those with more detailed class definitions—are incorporated.

\begin{figure}[!ht]
    \includegraphics[width=\linewidth]{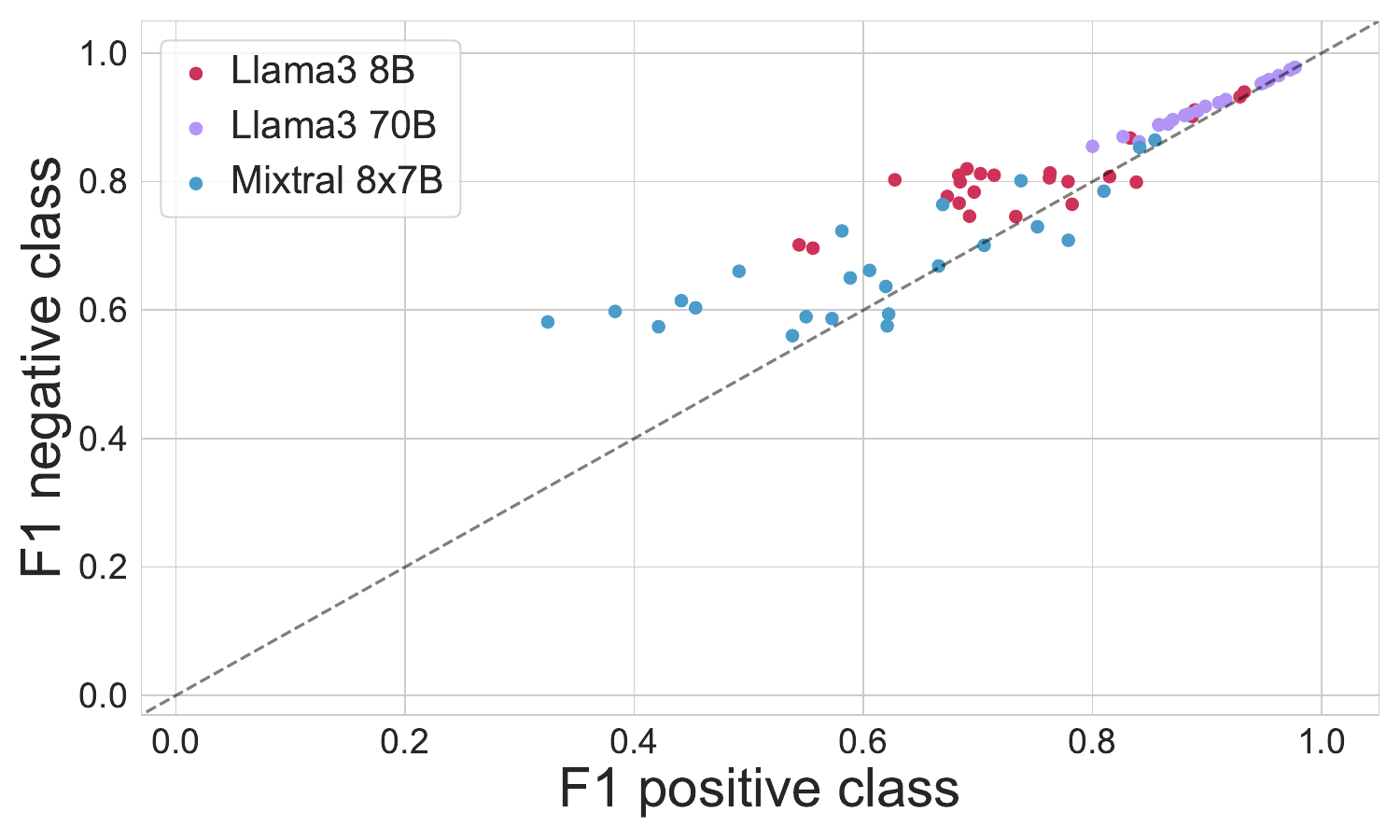}
    \caption{Task 1: Comparison between the F1 scores obtained by each model for each prompt variation with respect to both classes. The bisector is reported as a reference.}
    \label{fig:t1_f1_scatter}
\end{figure}

Figure~\ref{fig:t1_f1_scatter} highlights an approximately linear relationship between the F1 scores for the positive and negative classes, suggesting that the models maintain a balanced performance on this task. However, there is a slight tendency for less effective models, such as \llamaeight{} and \mixtral{}, to perform better on the negative class, possibly due to their smaller number of parameters.  

\begin{figure}[!ht]
    \centering
    \includegraphics[width=\linewidth]{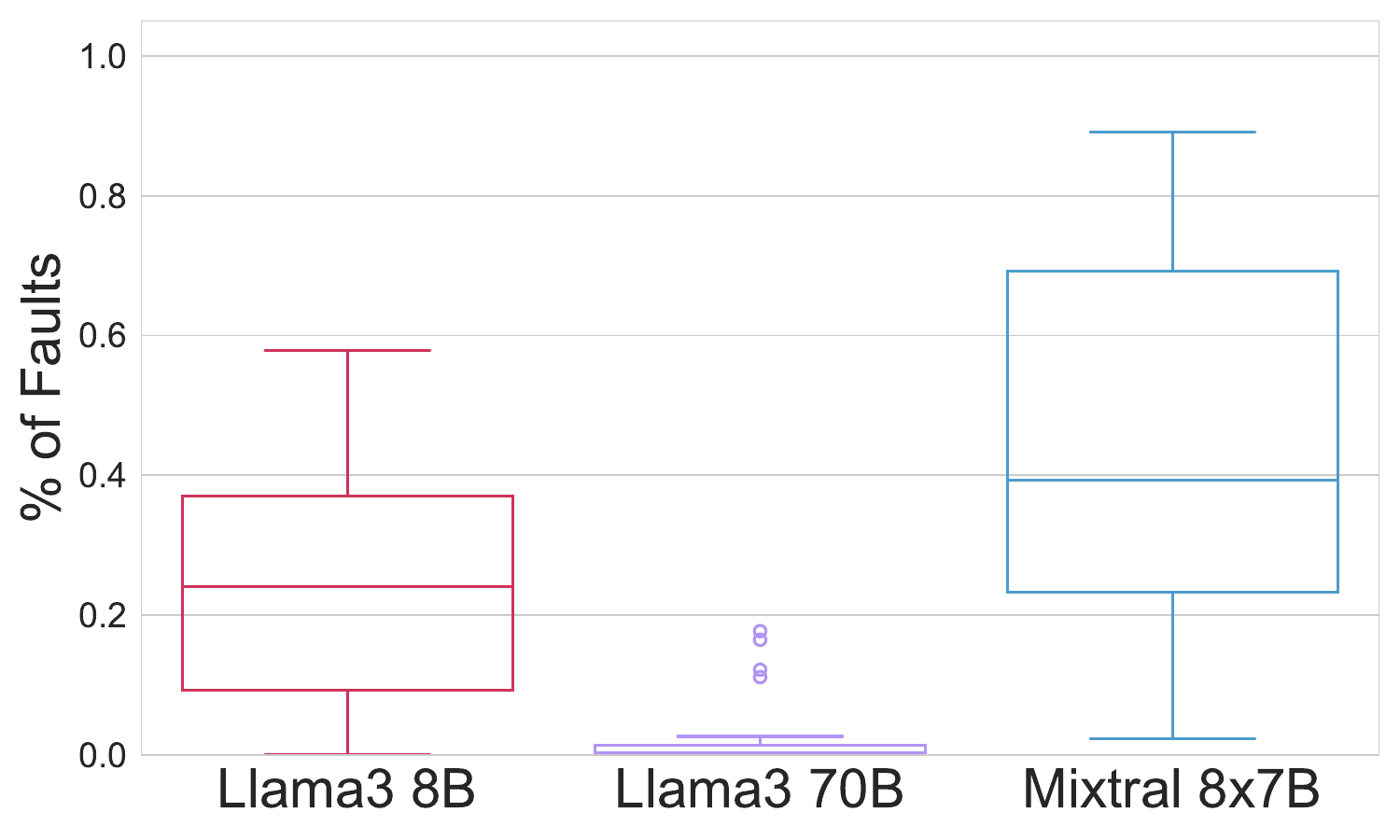}
    \caption{Task 1: Percentage of faults for each model. The median faults percentage for each model is: 0.241 (\llamaeight{}), 0.003 (\llamaseventy{}), 0.393 (\mixtral{})}
    \label{fig:t1_faults}
\end{figure}

Finally, \llamaseventy{} not only outperforms the other models in terms of F1 score but also generates the fewest faulty responses (i.e. instances where the model fails to provide a score or adhere to the expected output format), as illustrated in Fig.~\ref{fig:t1_faults}. In particular, it exhibits the lowest median percentage of faults: 0.003 compared to \llamaeight{}'s 0.241 and \mixtral{}'s 0.393.


\subsection*{Task 2: Providing an accurate verdict based on a fact-checking article}
In this task, the performance gap between the positive and negative classes is more pronounced, affecting not only the three evaluated models but also the fine-tuned \roberta{}.  

This discrepancy is primarily due to the class imbalance introduced by the 2024 samples in the dataset, which contain 470 false claims but only 30 true claims. This distribution contrasts with earlier years (2003–2013), where the dataset is balanced.  

Focusing on the results for the negative class, all three models perform relatively well compared to the baseline set by the fine-tuned \roberta{}, although none fully match its performance. Among them, \llamaseventy{} comes closest, achieving an F1 score of approximately 0.9, compared to \roberta{}'s 0.95.  

The performance gap widens even further for the positive class, which contains fewer samples. This is evident both in comparison to \roberta{}’s baseline and relative to the other models. \llamaseventy{} remains the strongest performer, consistently achieving scores above 0.6 across all configurations and coming closest to \roberta{}'s threshold.  

As observed in Task 1, no single prompting strategy proves to be the most effective across all models. Additionally, the rate of faulty responses increases for both \llamaeight{} and \llamaseventy{}, with the latter reaching nearly 30\% for some prompts. Meanwhile, \mixtral{} maintains a consistently high fault rate, similar to its performance in Task 1.

\begin{figure}[!ht]
    \centering
    \includegraphics[width=\linewidth]{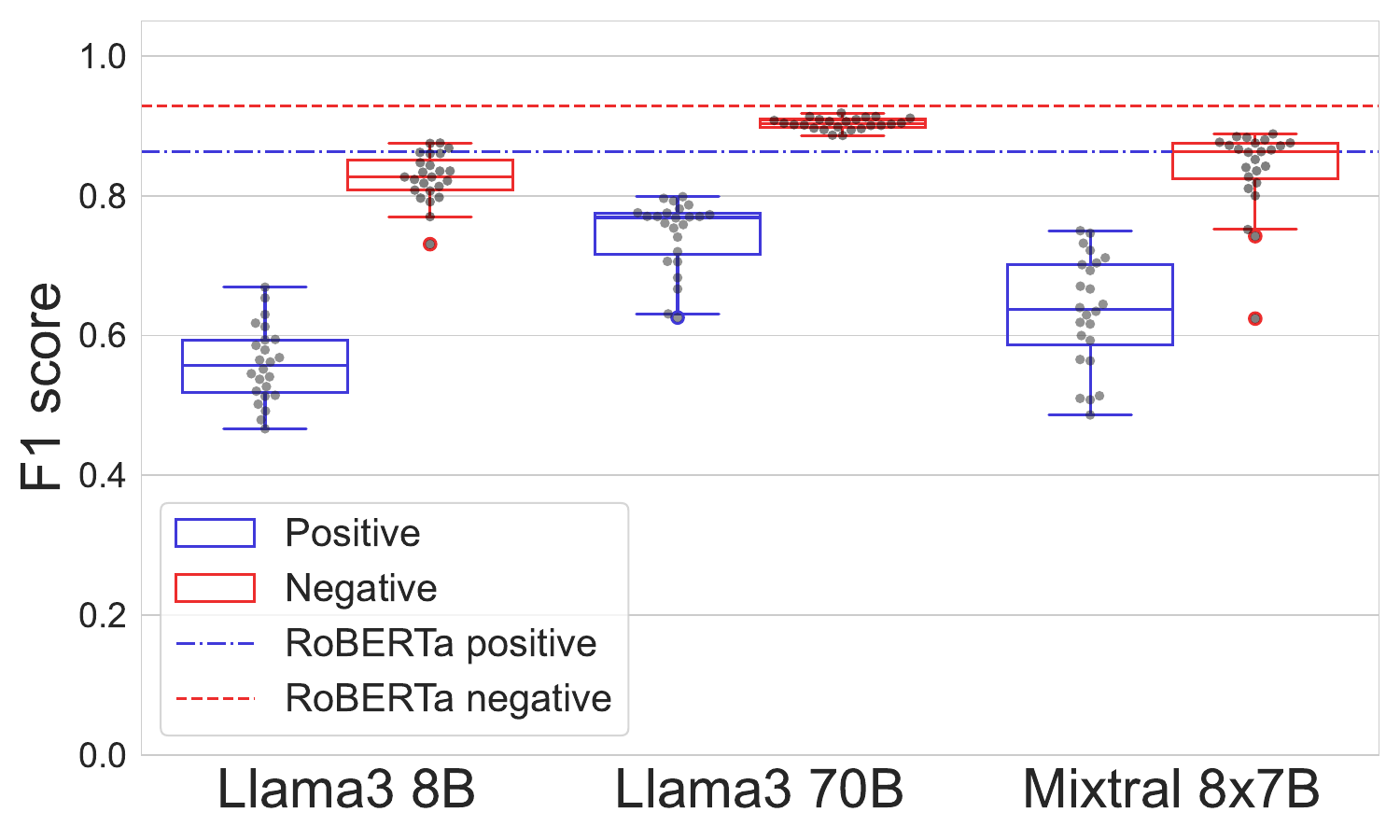}
    \caption{Task 2: Models' F1 scores computed for both classes. Fine-tuned \roberta{} is used as a reference baseline.}
    \label{fig:t2_f1_box}
\end{figure}

\begin{figure}[!ht]
    \centering
    \includegraphics[width=\linewidth]{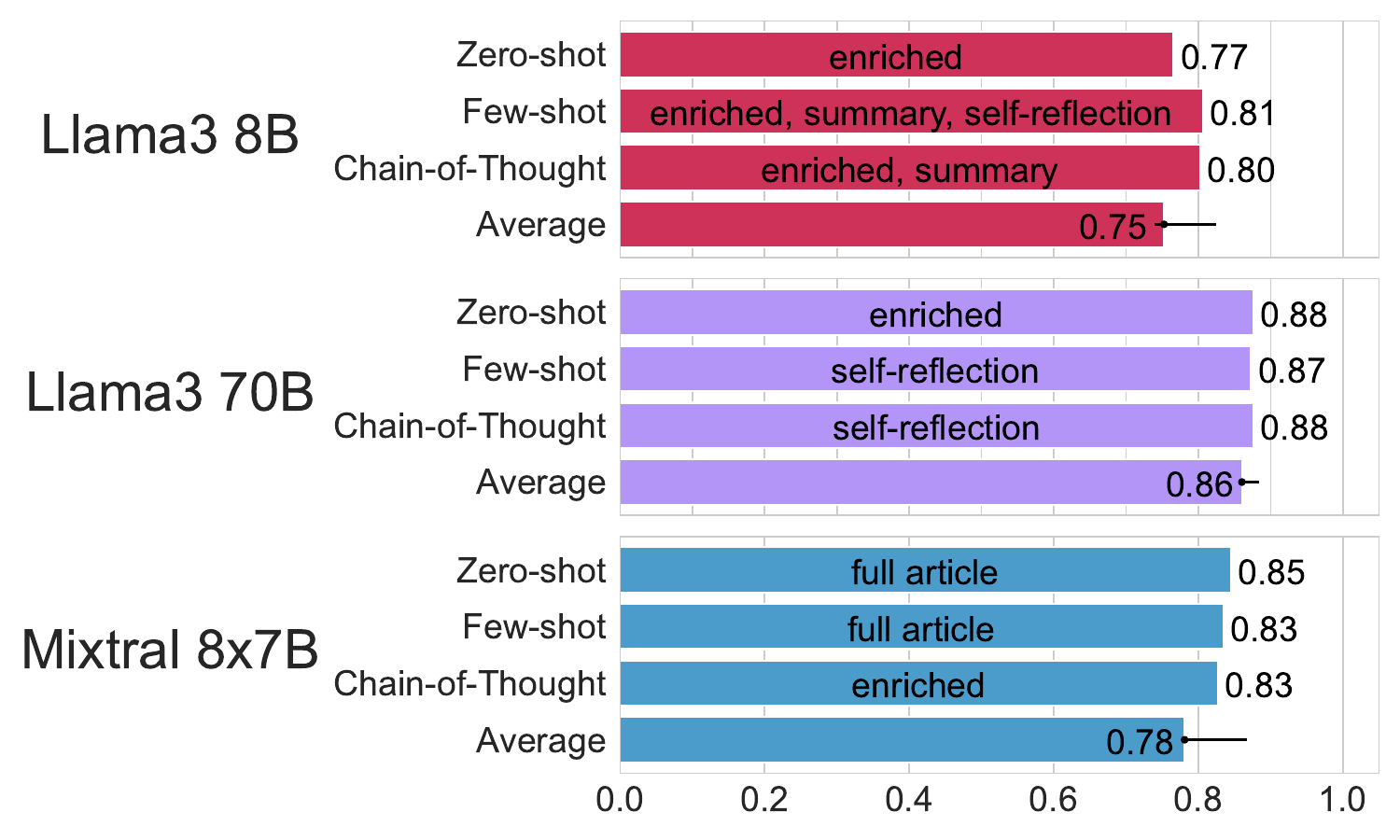}
    \caption{Task 2: For each model, for each prompt category (Zero-Shot, Few-Shot, and Chain-of-Tought) the best prompt's F1 score is reported. The average F1 score for each model is also reported with 0.95 confidence intervals.}
    \label{fig:t2_best_prompts}
\end{figure}

\begin{figure}[!ht]
    \centering
    \includegraphics[width=\linewidth]{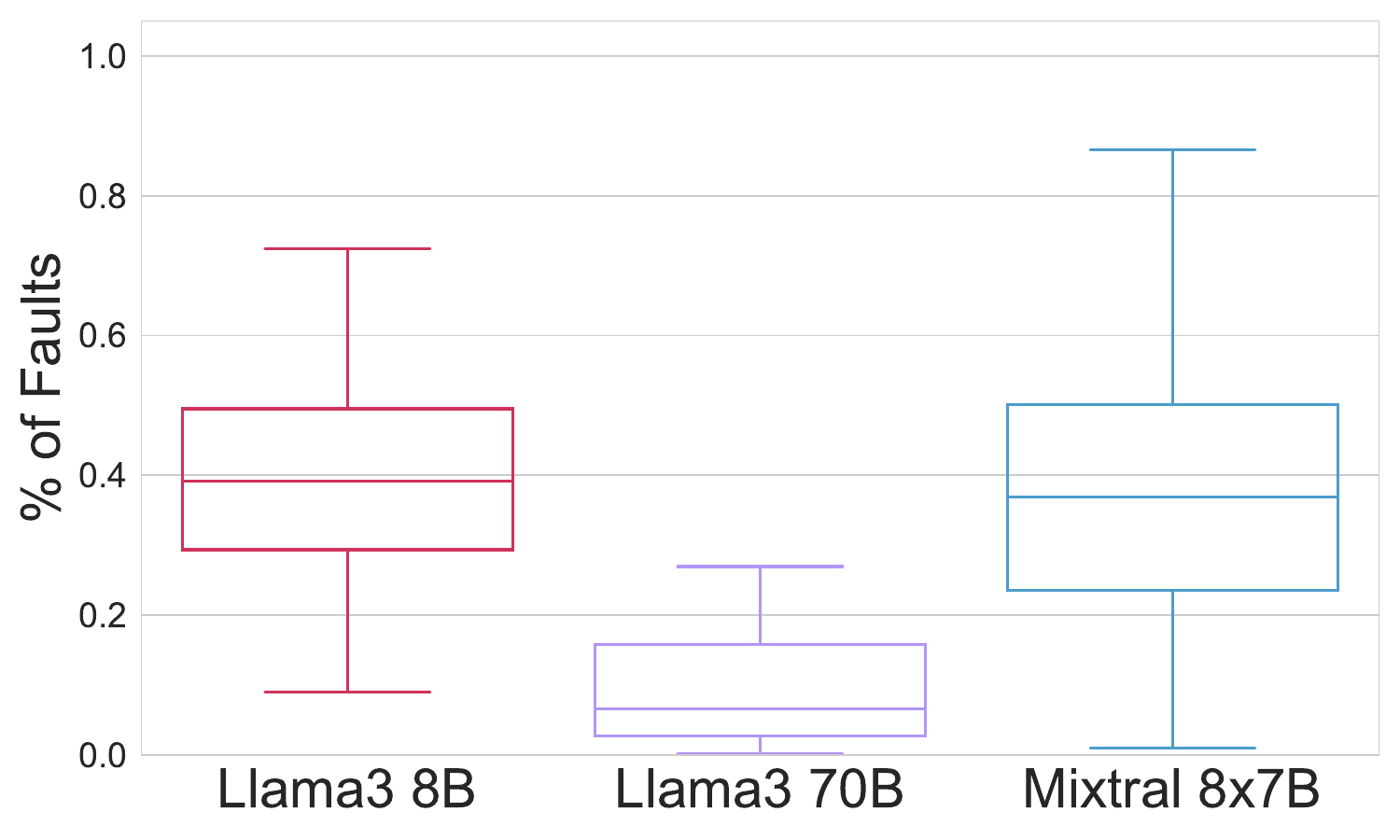}
    \caption{Task 2: The percentage of faults for each model. The median faults percentage for each model is: 0.392 (\llamaeight{}), 0.066 (\llamaseventy{}), 0.370 (\mixtral{})}
    \label{fig:t2_faults}
\end{figure}

\subsection*{Task 3: Fact-Checking the claim}

\begin{figure}[!ht]
    \centering
    \includegraphics[width=\linewidth]{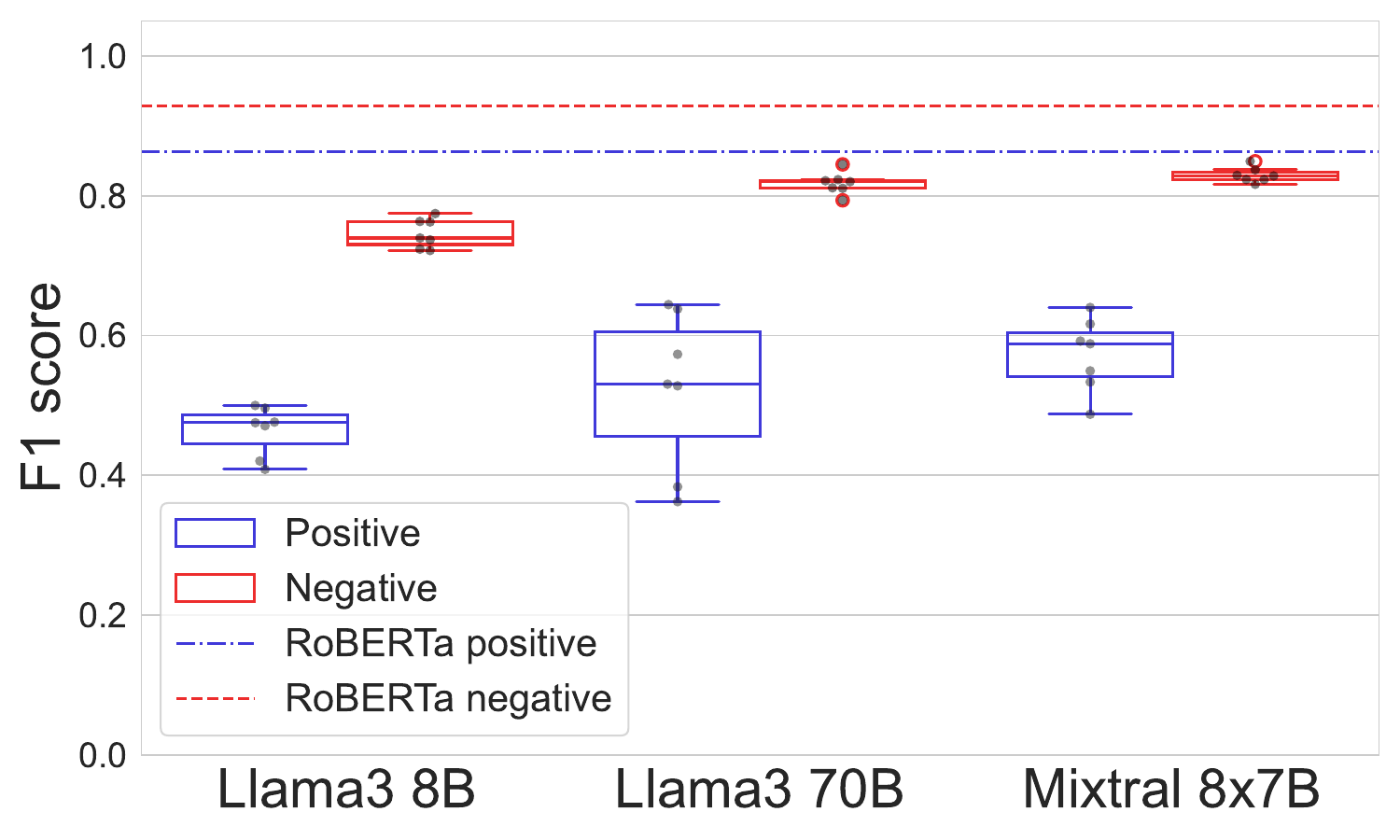}
    \caption{Task 3: Models' F1 scores computed for both classes. Fine-tuned \roberta{} is used as a reference baseline.}
    \label{fig:t3_f1_box}
\end{figure}

Figure~\ref{fig:t3_f1_box} shows that, under the current experimental settings, LLMs achieve high F1 scores for the negative class. However, they are consistently outperformed by the \roberta{} baseline. This suggests that while advanced models can sometimes perform well in classifying claims as \textbf{True} or \textbf{False}, simpler approaches may consistently yield better results. 

A notable trend in the results is the widening gap between the F1 scores of the positive and negative classes. Specifically, the highest F1 score achieved for the positive class is lower than the lowest F1 score observed for the negative class. This discrepancy highlights a significant challenge in the classification task. As was also the case in Task 2, this phenomenon can be attributed to dataset imbalance, which skews the model's ability to correctly predict both classes with equal proficiency.

\begin{figure}[!ht]
    \centering
    \includegraphics[width=\linewidth]{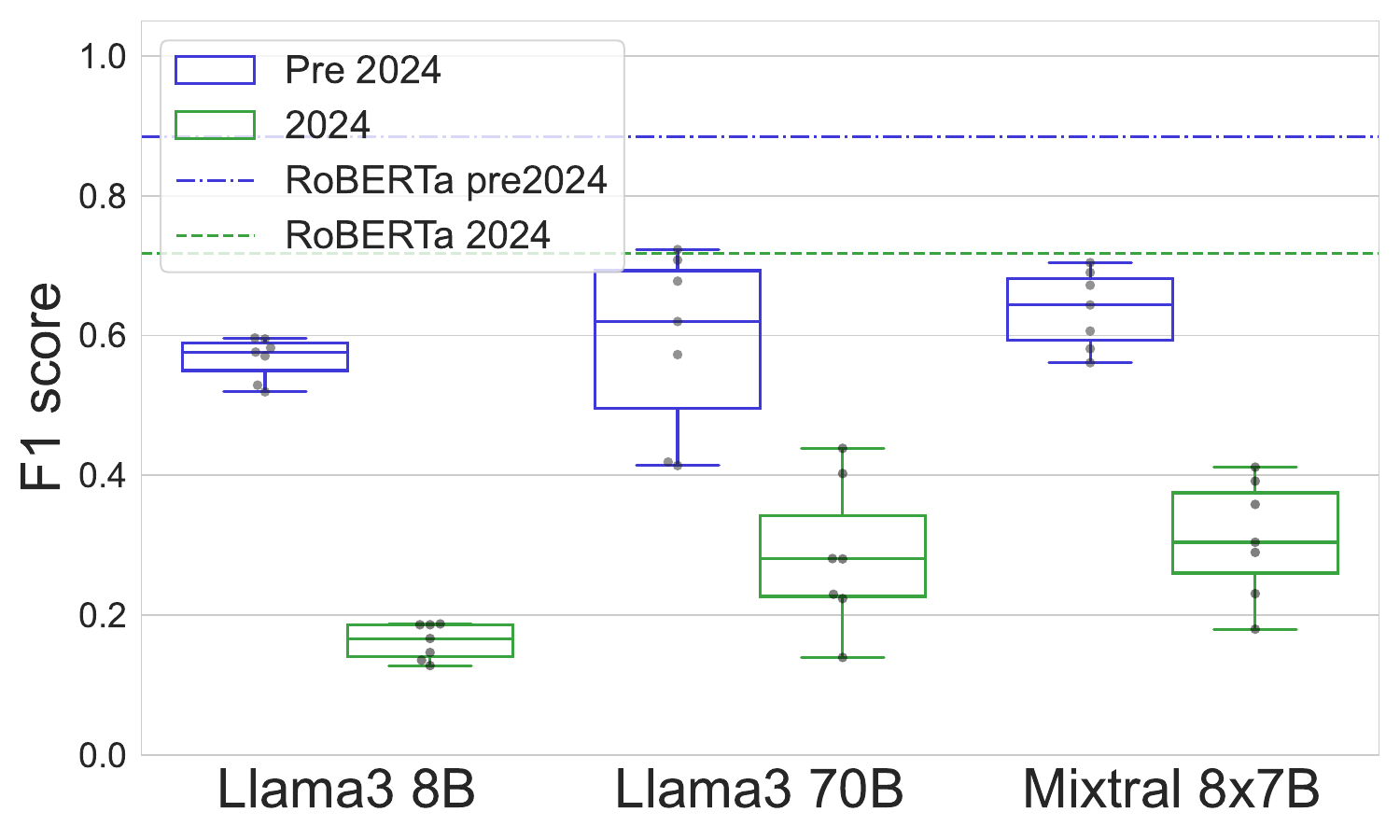}
    \caption{Task 3: Models' F1 scores computed for the positive class distinguishing between claims dated before 2024 (thus possibly included in the models' training dataset) and from 2024. Fine-tuned \roberta{} is used as a reference baseline.}
    \label{fig:t3_f1_box_pos}
\end{figure}

\begin{figure}[!ht]
    \centering
    \includegraphics[width=\linewidth]{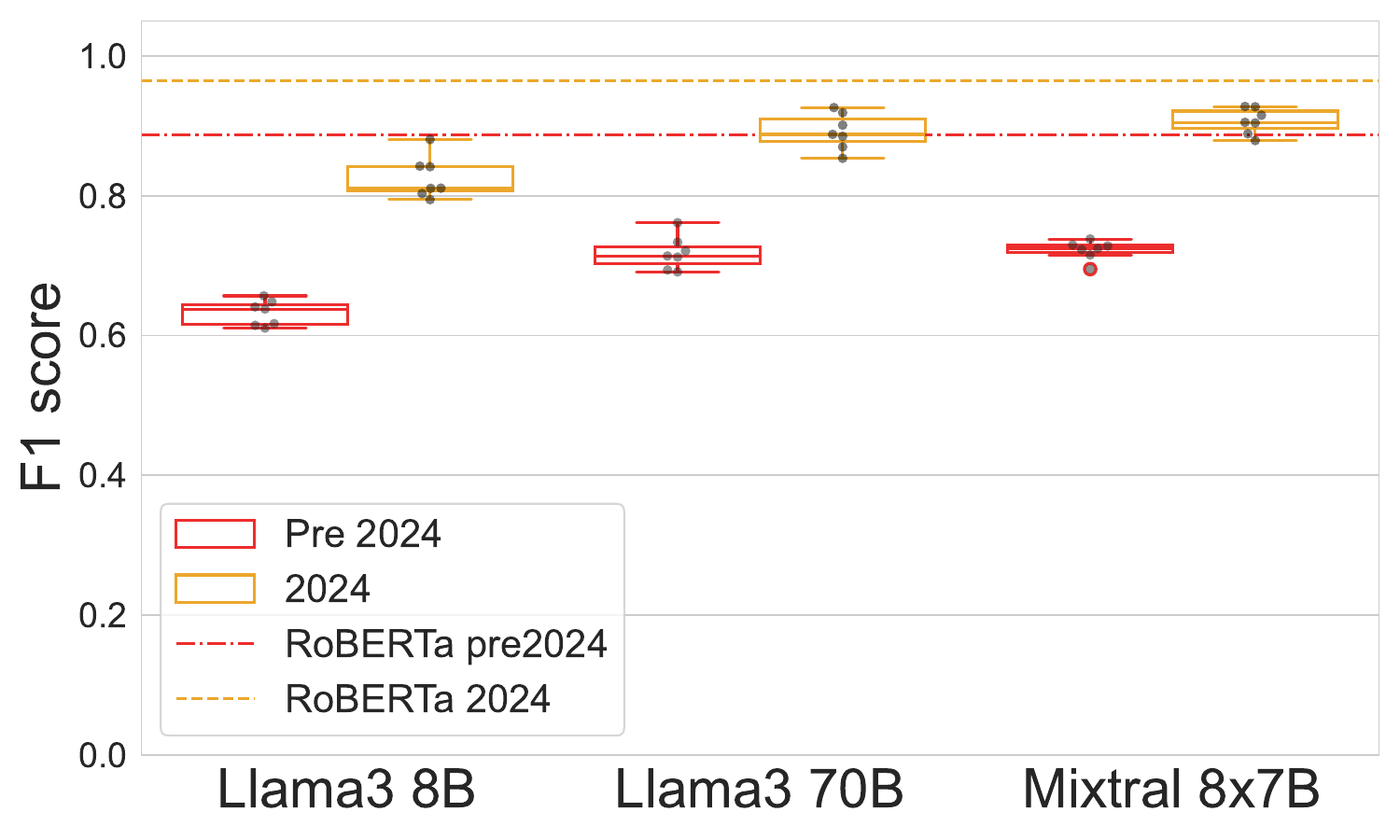}
    \caption{Task 3: Models' F1 scores computed for the negative class distinguishing between claims dated before 2024 (thus possibly included in the models' training dataset) and from 2024. Fine-tuned \roberta{} is used as a reference baseline.}
    \label{fig:t3_f1_box_neg}
\end{figure}

Further analysis reveals an intriguing temporal pattern when breaking down the performance by claim publication date. As depicted in Figure~\ref{fig:t3_f1_box_pos}, models exhibit substantially better performance on the positive class for claims that were published before 2024. Conversely, Figure~\ref{fig:t3_f1_box_neg} shows an opposite trend for the negative class: claims originating from 2024 tend to yield the highest performance. This suggests that temporal factors, possibly related to shifts in linguistic patterns, dataset composition, or the evolving nature of factual claims, may be influencing the model's ability to distinguish between true and false claims.

\begin{figure}[!ht]
    \centering
    \includegraphics[width=\linewidth]{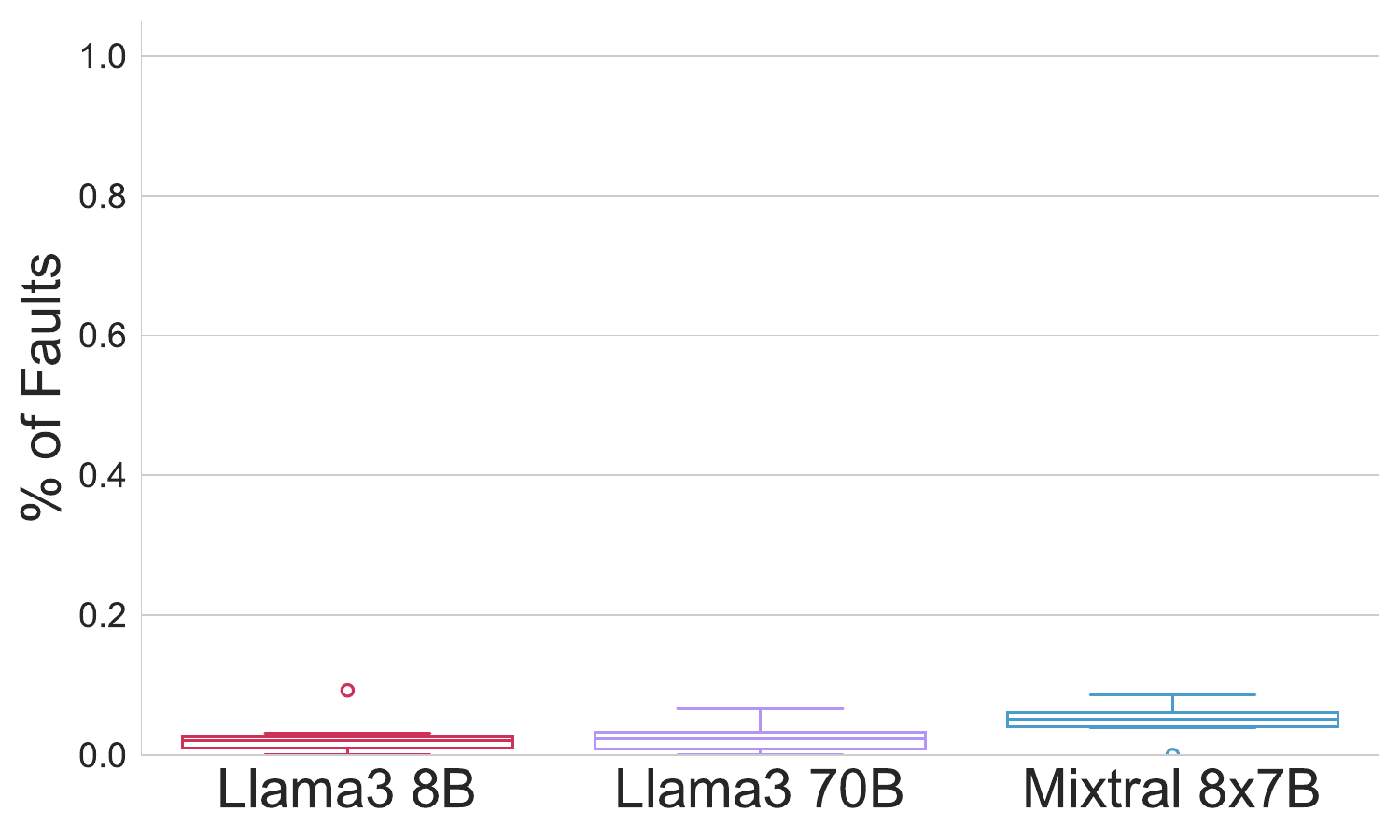}
    \caption{Task 3: The percentage of faults for each model. The median faults percentage for each model is: 0.02 (\llamaeight{}), 0.024 (\llamaseventy{}), 0.051 (\mixtral{})}
    \label{fig:t3_f1_faults}
\end{figure}

Unlike the previous tasks, one striking difference is the relatively low rate of faulty responses across all three models, as illustrated in Figure~\ref{fig:t3_f1_faults}. This suggests that, at least within the scope of this particular task, the models are more stable and less prone to generating erroneous classifications compared to their performance in earlier experiments.

\begin{figure}[!ht]
    \centering
    \includegraphics[width=\linewidth]{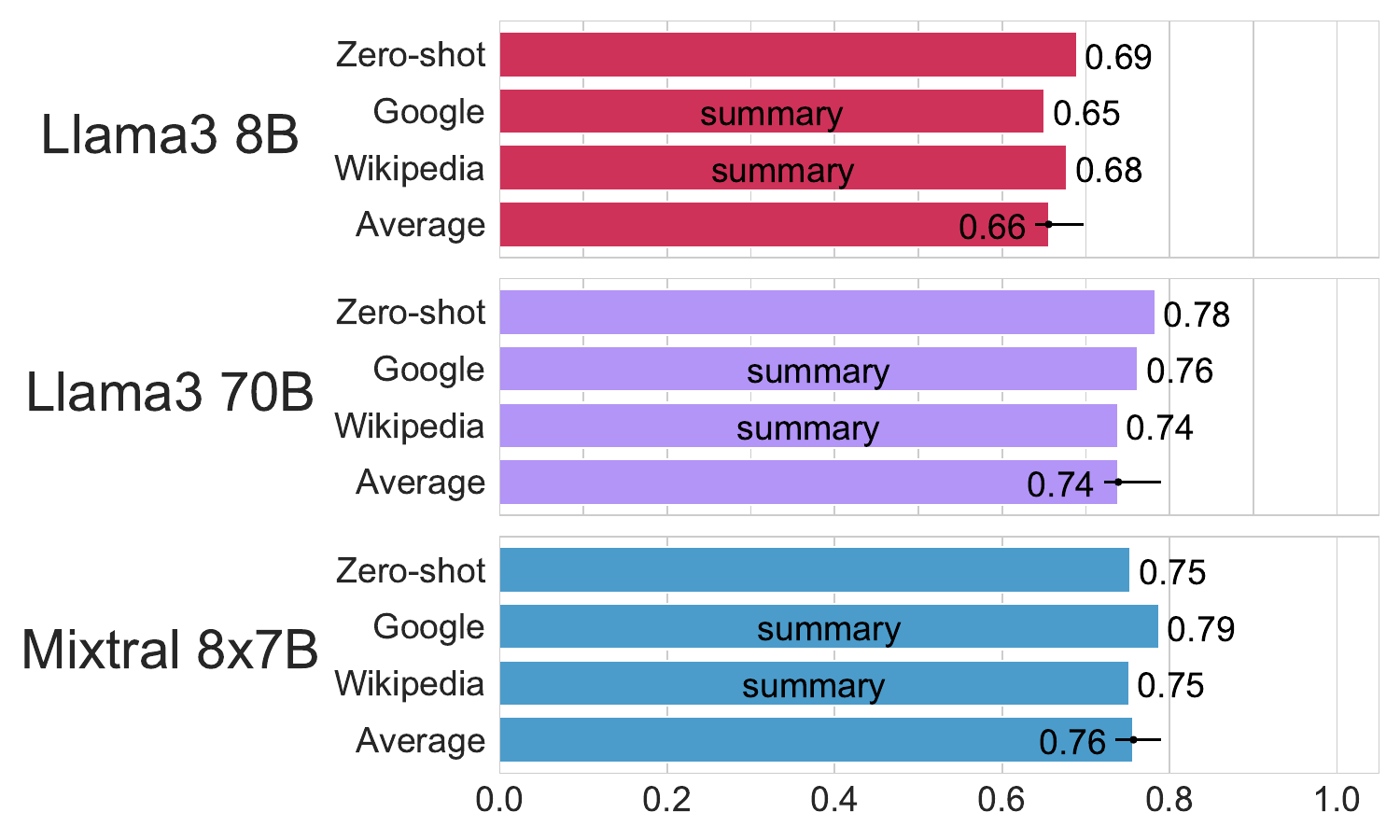}
    \caption{Task 3: For each model, for each prompt category (Zero-Shot, with Google contextual information, and with Wikipedia contextual information) the best prompt's F1 score is reported. The average F1 score for each model is also reported with 0.95 confidence intervals.}
    \label{fig:t3_best_prompts}
\end{figure}

Another interesting observation emerges when analyzing the impact of external content sources on model performance. Specifically, as shown in Figure~\ref{fig:t3_best_prompts}, when models are supplemented with information extracted from Google and Wikipedia, their classification accuracy improves. However, a crucial factor in this improvement appears to be the format in which the external information is presented. The results indicate that models perform significantly better when provided with a structured summary of the search results, rather than being fed entire web pages or isolated snippets. This suggests that concise, well-organized contextual information is more beneficial for improving model accuracy than raw, unstructured text.

\section{Conclusion}
Our study builds upon and expands existing research on the role of LLMs in fact-checking, offering new insights into their capabilities and limitations.

To systematically evaluate LLMs in this context, we designed our study around three key tasks. The first task assessed the models’ ability to recognize the semantic relationship between a claim and an article. The second task measured their performance in verifying a claim’s truthfulness when provided with a related fact-checking article. Finally, the third task explored the zero-shot fact-checking abilities of LLMs under varying levels of external knowledge support.

Our findings from the first task indicate that LLMs effectively identify the semantic connection between claims and articles, outperforming fine-tuned Small Language Models in this regard. This suggests a strong potential for LLMs in assisting fact-checking efforts. Additionally, we observed that prompt design plays a crucial role in model performance, highlighting the need for careful prompt engineering in fact-checking applications.

In the second task, we found that larger LLMs can accurately determine the veracity of claims when provided with a fact-checking article, particularly when evaluating fake news. However, in line with previous work~\cite{Quelle_2024}, their performance declines significantly when verifying true news, where simpler fine-tuned Small Language Models substantially outperform them.

Similarly, in the third task, all three LLMs underperformed compared to a fine-tuned \roberta{} model when asked to assess claim veracity—regardless of whether additional external knowledge (sourced from Google or Wikipedia) was incorporated. These results align with prior findings~\cite{HuBea24_BadActorGoodAdvisor}, which highlight the superior performance of fine-tuned Small Language Models over LLMs in fake news detection. This reinforces the idea that, while LLMs can be valuable tools in fact-checking, they are not yet reliable enough to fully automate the process.

Contrary to expectations from previous studies, introducing external knowledge did not enhance LLM performance. A possible explanation for this, as suggested by~\cite{WuJea24_FakeNewsInSheepsClothing}, is that fake and factual news often exhibit distinct writing styles, which may significantly hinder LLMs' ability to differentiate between them.

Overall, our study highlights both the promise and the challenges of using LLMs for fact-checking, emphasizing the need for further advancements before they can serve as a standalone solution.

\section*{Acknowledgments}
The authors are thankful to Sofia Mongardi for her support in the design of this manuscript.
The work in this paper was originally submitted as a Master Thesis titled: "Evaluating the Effectiveness of Open Large Language Models in Fact-checking Claims" written by Enrico Zuccolotto and supervised by Prof. Francesco Pierri. This paper is supported by PNRR-PE-AI FAIR project funded by the NextGeneration EU program.

\clearpage
\doublespacing
\bibliographystyle{IEEEtran}
\bibliography{bibliography}

\begin{thebibliography}{10}
\providecommand{\url}[1]{#1}
\csname url@samestyle\endcsname
\providecommand{\newblock}{\relax}
\providecommand{\bibinfo}[2]{#2}
\providecommand{\BIBentrySTDinterwordspacing}{\spaceskip=0pt\relax}
\providecommand{\BIBentryALTinterwordstretchfactor}{4}
\providecommand{\BIBentryALTinterwordspacing}{\spaceskip=\fontdimen2\font plus
\BIBentryALTinterwordstretchfactor\fontdimen3\font minus
  \fontdimen4\font\relax}
\providecommand{\BIBforeignlanguage}[2]{{%
\expandafter\ifx\csname l@#1\endcsname\relax
\typeout{** WARNING: IEEEtran.bst: No hyphenation pattern has been}%
\typeout{** loaded for the language `#1'. Using the pattern for}%
\typeout{** the default language instead.}%
\else
\language=\csname l@#1\endcsname
\fi
#2}}
\providecommand{\BIBdecl}{\relax}
\BIBdecl

\bibitem{DelVicarioMea16_SpreadingMisinformationOnline}
\BIBentryALTinterwordspacing
M.~Del~Vicario, A.~Bessi, F.~Zollo, F.~Petroni, A.~Scala, G.~Caldarelli, H.~E.
  Stanley, and W.~Quattrociocchi, ``The spreading of misinformation online,''
  \emph{Proceedings of the National Academy of Sciences}, vol. 113, no.~3, p.
  554–559, Jan. 2016. [Online]. Available:
  \url{https://www.pnas.org/doi/full/10.1073/pnas.1517441113}
\BIBentrySTDinterwordspacing

\bibitem{FakeNewsInfluencingElections}
\BIBentryALTinterwordspacing
H.~Allcott and M.~Gentzkow, ``Social media and fake news in the 2016
  election,'' \emph{Journal of Economic Perspectives}, vol.~31, no.~2, p.
  211–36, May 2017. [Online]. Available:
  \url{https://www.aeaweb.org/articles?id=10.1257/jep.31.2.211}
\BIBentrySTDinterwordspacing

\bibitem{BarDea24_SystematicDiscrepanciesDelivery}
\BIBentryALTinterwordspacing
D.~Bär, F.~Pierri, G.~De~Francisci~Morales, and S.~Feuerriegel, ``Systematic
  discrepancies in the delivery of political ads on facebook and instagram,''
  \emph{PNAS Nexus}, vol.~3, no.~7, p. pgae247, Jul. 2024a. [Online].
  Available: \url{https://doi.org/10.1093/pnasnexus/pgae247}
\BIBentrySTDinterwordspacing

\bibitem{LackOfTrust}
\BIBentryALTinterwordspacing
K.~Fink, ``The biggest challenge facing journalism: A lack of trust,''
  \emph{Journalism}, vol.~20, no.~1, pp. 40--43, 2019. [Online]. Available:
  \url{https://doi.org/10.1177/1464884918807069}
\BIBentrySTDinterwordspacing

\bibitem{spivak2010fact}
C.~Spivak, ``The fact-checking explosion: In a bitter political landscape
  marked by rampant allegations of questionable credibility, more and more news
  outlets are launching truth-squad operations,'' \emph{American Journalism
  Review}, vol.~32, no.~4, pp. 38--44, 2010.

\bibitem{Lim2018}
C.~Lim, ``Checking how fact-checkers check,'' \emph{Research \& Politics},
  vol.~5, no.~3, 2018.

\bibitem{WarrenGea25p_ShowMeTheWork}
\BIBentryALTinterwordspacing
G.~Warren, I.~Shklovski, and I.~Augenstein, ``Show me the work:
  Fact-checkers’ requirements for explainable automated fact-checking,'' Feb.
  2025, arXiv:2502.09083 [cs]. [Online]. Available:
  \url{http://arxiv.org/abs/2502.09083}
\BIBentrySTDinterwordspacing

\bibitem{Guo2022}
Z.~Guo, M.~Schlichtkrull, and A.~Vlachos, ``A survey on automated
  fact-checking,'' \emph{Transactions of the Association for Computational
  Linguistics}, vol.~10, pp. 178--206, 2022.

\bibitem{HU2022133}
\BIBentryALTinterwordspacing
L.~Hu, S.~Wei, Z.~Zhao, and B.~Wu, ``Deep learning for fake news detection: A
  comprehensive survey,'' \emph{AI Open}, vol.~3, pp. 133--155, 2022. [Online].
  Available:
  \url{https://www.sciencedirect.com/science/article/pii/S2666651022000134}
\BIBentrySTDinterwordspacing

\bibitem{GNN_all}
\BIBentryALTinterwordspacing
Y.~Wang, S.~Qian, J.~Hu, Q.~Fang, and C.~Xu, ``Fake news detection via
  knowledge-driven multimodal graph convolutional networks,'' in
  \emph{Proceedings of the 2020 International Conference on Multimedia
  Retrieval}, ser. ICMR '20.\hskip 1em plus 0.5em minus 0.4em\relax New York,
  NY, USA: Association for Computing Machinery, 2020, p. 540–547. [Online].
  Available: \url{https://doi.org/10.1145/3372278.3390713}
\BIBentrySTDinterwordspacing

\bibitem{LIN2022111}
\BIBentryALTinterwordspacing
T.~Lin, Y.~Wang, X.~Liu, and X.~Qiu, ``A survey of transformers,'' \emph{AI
  Open}, vol.~3, pp. 111--132, 2022. [Online]. Available:
  \url{https://www.sciencedirect.com/science/article/pii/S2666651022000146}
\BIBentrySTDinterwordspacing

\bibitem{PLMs}
A.~Vaswani, N.~Shazeer, N.~Parmar, J.~Uszkoreit, L.~Jones, A.~N. Gomez,
  {\L}.~Kaiser, and I.~Polosukhin, ``Attention is all you need,''
  \emph{Advances in neural information processing systems}, vol.~30, 2017.

\bibitem{hadi2023survey}
M.~U. Hadi, R.~Qureshi, A.~Shah, M.~Irfan, A.~Zafar, M.~B. Shaikh, N.~Akhtar,
  J.~Wu, S.~Mirjalili \emph{et~al.}, ``A survey on large language models:
  Applications, challenges, limitations, and practical usage,'' \emph{Authorea
  Preprints}, 2023.

\bibitem{FontanaNea24p_NicerThanHumans}
\BIBentryALTinterwordspacing
N.~Fontana, F.~Pierri, and L.~M. Aiello, ``Nicer than humans: How do large
  language models behave in the prisoner’s dilemma?'' Sep. 2024,
  arXiv:2406.13605. [Online]. Available: \url{http://arxiv.org/abs/2406.13605}
\BIBentrySTDinterwordspacing

\bibitem{PapageorgiouEea24_SurveyUseLarge}
\BIBentryALTinterwordspacing
E.~Papageorgiou, C.~Chronis, I.~Varlamis, and Y.~Himeur,
  ``\BIBforeignlanguage{en}{A survey on the use of large language models (llms)
  in fake news},'' \emph{\BIBforeignlanguage{en}{Future Internet}}, vol.~16,
  no.~88, p. 298, Aug. 2024. [Online]. Available:
  \url{https://www.mdpi.com/1999-5903/16/8/298}
\BIBentrySTDinterwordspacing

\bibitem{SanuEea24_LimitationsLargeLanguage}
\BIBentryALTinterwordspacing
E.~Sanu, T.~K. Amudaa, P.~Bhat, G.~Dinesh, A.~U. Kumar~Chate, and R.~K. P,
  ``Limitations of large language models,'' in \emph{2024 8th International
  Conference on Computational System and Information Technology for Sustainable
  Solutions (CSITSS)}, Nov. 2024, p. 1–6. [Online]. Available:
  \url{https://ieeexplore.ieee.org/abstract/document/10817070}
\BIBentrySTDinterwordspacing

\bibitem{Quelle_2024}
\BIBentryALTinterwordspacing
D.~Quelle and A.~Bovet, ``The perils and promises of fact-checking with large
  language models,'' \emph{Frontiers in Artificial Intelligence}, vol.~7, Feb.
  2024. [Online]. Available: \url{http://dx.doi.org/10.3389/frai.2024.1341697}
\BIBentrySTDinterwordspacing

\bibitem{PierriFea19_FalseNewsOnSocialMedia}
\BIBentryALTinterwordspacing
F.~Pierri and S.~Ceri, ``False news on social media: A data-driven survey,''
  \emph{SIGMOD Rec.}, vol.~48, no.~2, p. 18–27, Dec. 2019. [Online].
  Available: \url{https://dl.acm.org/doi/10.1145/3377330.3377334}
\BIBentrySTDinterwordspacing

\bibitem{NasirJea21_FakeNewsDetection}
\BIBentryALTinterwordspacing
J.~A. Nasir, O.~S. Khan, and I.~Varlamis, ``Fake news detection: A hybrid
  cnn-rnn based deep learning approach,'' \emph{International Journal of
  Information Management Data Insights}, vol.~1, no.~1, p. 100007, Apr. 2021b.
  [Online]. Available:
  \url{https://www.sciencedirect.com/science/article/pii/S2667096820300070}
\BIBentrySTDinterwordspacing

\bibitem{MewadaAea24_CIPF}
\BIBentryALTinterwordspacing
A.~Mewada and R.~K. Dewang, ``\BIBforeignlanguage{en}{Cipf: Identifying fake
  profiles on social media using a cnn-based communal influence propagation
  framework},'' \emph{\BIBforeignlanguage{en}{Multimedia Tools and
  Applications}}, vol.~83, no.~10, p. 29419–29454, Mar. 2024a. [Online].
  Available: \url{https://doi.org/10.1007/s11042-023-16685-z}
\BIBentrySTDinterwordspacing

\bibitem{SuJea24_AdaptingFakeNews}
\BIBentryALTinterwordspacing
J.~Su, C.~Cardie, and P.~Nakov, ``Adapting fake news detection to the era of
  large language models,'' in \emph{Findings of the Association for
  Computational Linguistics: NAACL 2024}, K.~Duh, H.~Gomez, and S.~Bethard,
  Eds.\hskip 1em plus 0.5em minus 0.4em\relax Mexico City, Mexico: Association
  for Computational Linguistics, Jun. 2024, p. 1473–1490. [Online].
  Available: \url{https://aclanthology.org/2024.findings-naacl.95/}
\BIBentrySTDinterwordspacing

\bibitem{ZhangXea21_MiningDualEmotion}
\BIBentryALTinterwordspacing
X.~Zhang, J.~Cao, X.~Li, Q.~Sheng, L.~Zhong, and K.~Shu, ``Mining dual emotion
  for fake news detection,'' in \emph{Proceedings of the Web Conference 2021},
  ser. WWW ’21.\hskip 1em plus 0.5em minus 0.4em\relax New York, NY, USA:
  Association for Computing Machinery, Jun. 2021, p. 3465–3476. [Online].
  Available: \url{https://dl.acm.org/doi/10.1145/3442381.3450004}
\BIBentrySTDinterwordspacing

\bibitem{ZhangXea24_ReinforcementRetrievalLeveraging}
\BIBentryALTinterwordspacing
X.~Zhang and W.~Gao, ``Reinforcement retrieval leveraging fine-grained feedback
  for fact checking news claims with black-box llm,'' in \emph{Proceedings of
  the 2024 Joint International Conference on Computational Linguistics,
  Language Resources and Evaluation (LREC-COLING 2024)}, N.~Calzolari, M.-Y.
  Kan, V.~Hoste, A.~Lenci, S.~Sakti, and N.~Xue, Eds.\hskip 1em plus 0.5em
  minus 0.4em\relax Torino, Italia: ELRA and ICCL, May 2024, p. 13861–13873.
  [Online]. Available: \url{https://aclanthology.org/2024.lrec-main.1209/}
\BIBentrySTDinterwordspacing

\bibitem{ZhangXea23_LLMbasedFactVerification}
\BIBentryALTinterwordspacing
------, ``Towards llm-based fact verification on news claims with a
  hierarchical step-by-step prompting method,'' in \emph{Proceedings of the
  13th International Joint Conference on Natural Language Processing and the
  3rd Conference of the Asia-Pacific Chapter of the Association for
  Computational Linguistics (Volume 1: Long Papers)}, J.~C. Park, Y.~Arase,
  B.~Hu, W.~Lu, D.~Wijaya, A.~Purwarianti, and A.~A. Krisnadhi, Eds.\hskip 1em
  plus 0.5em minus 0.4em\relax Nusa Dua, Bali: Association for Computational
  Linguistics, Nov. 2023a, p. 996–1011. [Online]. Available:
  \url{https://aclanthology.org/2023.ijcnlp-main.64/}
\BIBentrySTDinterwordspacing

\bibitem{NogaraGea24p_ToxicBias}
\BIBentryALTinterwordspacing
G.~Nogara, F.~Pierri, S.~Cresci, L.~Luceri, P.~Törnberg, and S.~Giordano,
  ``Toxic bias: Perspective api misreads german as more toxic,'' Jul. 2024c,
  arXiv:2312.12651 [cs]. [Online]. Available:
  \url{http://arxiv.org/abs/2312.12651}
\BIBentrySTDinterwordspacing

\bibitem{LiuGea25p_ComparingDiversityNegativityAndStereotypesInChinese-languageAITechnologies}
\BIBentryALTinterwordspacing
G.~Liu, C.~A. Bono, and F.~Pierri, ``Comparing diversity, negativity, and
  stereotypes in chinese-language ai technologies: an investigation of baidu,
  ernie and qwen,'' Feb. 2025b, arXiv:2408.15696 [cs]. [Online]. Available:
  \url{http://arxiv.org/abs/2408.15696}
\BIBentrySTDinterwordspacing

\bibitem{AugensteinIea24_FactualityChallengesEra}
\BIBentryALTinterwordspacing
I.~Augenstein, T.~Baldwin, M.~Cha, T.~Chakraborty, G.~L. Ciampaglia, D.~Corney,
  R.~DiResta, E.~Ferrara, S.~Hale, A.~Halevy, E.~Hovy, H.~Ji, F.~Menczer,
  R.~Miguez, P.~Nakov, D.~Scheufele, S.~Sharma, and G.~Zagni,
  ``\BIBforeignlanguage{en}{Factuality challenges in the era of large language
  models and opportunities for fact-checking},''
  \emph{\BIBforeignlanguage{en}{Nature Machine Intelligence}}, vol.~6, no.~8,
  p. 852–863, Aug. 2024. [Online]. Available:
  \url{https://www.nature.com/articles/s42256-024-00881-z}
\BIBentrySTDinterwordspacing

\bibitem{YaoSea23_ReAct}
\BIBentryALTinterwordspacing
S.~Yao, J.~Zhao, D.~Yu, N.~Du, I.~Shafran, K.~Narasimhan, and Y.~Cao,
  ``\BIBforeignlanguage{en}{React: Synergizing reasoning and acting in language
  models},'' \emph{\BIBforeignlanguage{en}{International Conference on Learning
  Representations (ICLR)}}, Jan. 2023. [Online]. Available:
  \url{https://par.nsf.gov/biblio/10451467-react-synergizing-reasoning-acting-language-models}
\BIBentrySTDinterwordspacing

\bibitem{MinSea23_FActScore}
\BIBentryALTinterwordspacing
S.~Min, K.~Krishna, X.~Lyu, M.~Lewis, W.-t. Yih, P.~Koh, M.~Iyyer,
  L.~Zettlemoyer, and H.~Hajishirzi, ``Factscore: Fine-grained atomic
  evaluation of factual precision in long form text generation,'' in
  \emph{Proceedings of the 2023 Conference on Empirical Methods in Natural
  Language Processing}, H.~Bouamor, J.~Pino, and K.~Bali, Eds.\hskip 1em plus
  0.5em minus 0.4em\relax Singapore: Association for Computational Linguistics,
  Dec. 2023, p. 12076–12100. [Online]. Available:
  \url{https://aclanthology.org/2023.emnlp-main.741/}
\BIBentrySTDinterwordspacing

\bibitem{ChernIea23p_FacTool}
\BIBentryALTinterwordspacing
I.-C. Chern, S.~Chern, S.~Chen, W.~Yuan, K.~Feng, C.~Zhou, J.~He, G.~Neubig,
  and P.~Liu, ``Factool: Factuality detection in generative ai -- a tool
  augmented framework for multi-task and multi-domain scenarios,'' jul 2023,
  arXiv:2307.13528 [cs]. [Online]. Available:
  \url{http://arxiv.org/abs/2307.13528}
\BIBentrySTDinterwordspacing

\bibitem{SchickTea23_Toolformer}
\BIBentryALTinterwordspacing
T.~Schick, J.~Dwivedi-Yu, R.~Dessi, R.~Raileanu, M.~Lomeli, E.~Hambro,
  L.~Zettlemoyer, N.~Cancedda, and T.~Scialom,
  ``\BIBforeignlanguage{en}{Toolformer: Language models can teach themselves to
  use tools},'' \emph{\BIBforeignlanguage{en}{Advances in Neural Information
  Processing Systems}}, vol.~36, p. 68539–68551, Dec. 2023. [Online].
  Available:
  \url{https://proceedings.neurips.cc/paper_files/paper/2023/hash/d842425e4bf79ba039352da0f658a906-Abstract-Conference.html}
\BIBentrySTDinterwordspacing

\bibitem{WangYea24_Factcheck-Bench}
\BIBentryALTinterwordspacing
Y.~Wang, R.~Gangi~Reddy, Z.~M. Mujahid, A.~Arora, A.~Rubashevskii, J.~Geng,
  O.~Mohammed~Afzal, L.~Pan, N.~Borenstein, A.~Pillai, I.~Augenstein,
  I.~Gurevych, and P.~Nakov, ``Factcheck-bench: Fine-grained evaluation
  benchmark for automatic fact-checkers,'' in \emph{Findings of the Association
  for Computational Linguistics: EMNLP 2024}, Y.~Al-Onaizan, M.~Bansal, and
  Y.-N. Chen, Eds.\hskip 1em plus 0.5em minus 0.4em\relax Miami, Florida, USA:
  Association for Computational Linguistics, Nov. 2024, p. 14199–14230.
  [Online]. Available: \url{https://aclanthology.org/2024.findings-emnlp.830}
\BIBentrySTDinterwordspacing

\bibitem{DuYea24_ImprovingFactualityReasoning}
Y.~Du, S.~Li, A.~Torralba, J.~B. Tenenbaum, and I.~Mordatch, ``Improving
  factuality and reasoning in language models through multiagent debate,'' in
  \emph{Proceedings of the 41st International Conference on Machine Learning},
  ser. ICML’24, vol. 235.\hskip 1em plus 0.5em minus 0.4em\relax Vienna,
  Austria: JMLR.org, Jul. 2024, p. 11733–11763.

\bibitem{kim2024llmsproducefaithfulexplanations}
\BIBentryALTinterwordspacing
K.~Kim, S.~Lee, K.-H. Huang, H.~P. Chan, M.~Li, and H.~Ji, ``Can llms produce
  faithful explanations for fact-checking? towards faithful explainable
  fact-checking via multi-agent debate,'' 2024. [Online]. Available:
  \url{https://arxiv.org/abs/2402.07401}
\BIBentrySTDinterwordspacing

\bibitem{zhou2024correctingmisinformationsocialmedia}
\BIBentryALTinterwordspacing
X.~Zhou, A.~Sharma, A.~X. Zhang, and T.~Althoff, ``Correcting misinformation on
  social media with a large language model,'' 2024. [Online]. Available:
  \url{https://arxiv.org/abs/2403.11169}
\BIBentrySTDinterwordspacing

\bibitem{ZhaoRea23_Verify-and-Edit}
\BIBentryALTinterwordspacing
R.~Zhao, X.~Li, S.~Joty, C.~Qin, and L.~Bing, ``Verify-and-edit: A
  knowledge-enhanced chain-of-thought framework,'' in \emph{Proceedings of the
  61st Annual Meeting of the Association for Computational Linguistics (Volume
  1: Long Papers)}, A.~Rogers, J.~Boyd-Graber, and N.~Okazaki, Eds.\hskip 1em
  plus 0.5em minus 0.4em\relax Toronto, Canada: Association for Computational
  Linguistics, Jul. 2023, p. 5823–5840. [Online]. Available:
  \url{https://aclanthology.org/2023.acl-long.320/}
\BIBentrySTDinterwordspacing

\bibitem{DhuliawalaSea24_ChainofVerificationReducesHallucination}
\BIBentryALTinterwordspacing
S.~Dhuliawala, M.~Komeili, J.~Xu, R.~Raileanu, X.~Li, A.~Celikyilmaz, and
  J.~Weston, ``Chain-of-verification reduces hallucination in large language
  models,'' in \emph{Findings of the Association for Computational Linguistics:
  ACL 2024}, L.-W. Ku, A.~Martins, and V.~Srikumar, Eds.\hskip 1em plus 0.5em
  minus 0.4em\relax Bangkok, Thailand: Association for Computational
  Linguistics, Aug. 2024, p. 3563–3578. [Online]. Available:
  \url{https://aclanthology.org/2024.findings-acl.212/}
\BIBentrySTDinterwordspacing

\bibitem{LiMea24_Self-Checker}
\BIBentryALTinterwordspacing
M.~Li, B.~Peng, M.~Galley, J.~Gao, and Z.~Zhang, ``Self-checker: Plug-and-play
  modules for fact-checking with large language models,'' in \emph{Findings of
  the Association for Computational Linguistics: NAACL 2024}, K.~Duh, H.~Gomez,
  and S.~Bethard, Eds.\hskip 1em plus 0.5em minus 0.4em\relax Mexico City,
  Mexico: Association for Computational Linguistics, Jun. 2024, p. 163–181.
  [Online]. Available: \url{https://aclanthology.org/2024.findings-naacl.12/}
\BIBentrySTDinterwordspacing

\bibitem{HuBea24_BadActorGoodAdvisor}
\BIBentryALTinterwordspacing
B.~Hu, Q.~Sheng, J.~Cao, Y.~Shi, Y.~Li, D.~Wang, and P.~Qi,
  ``\BIBforeignlanguage{en}{Bad actor, good advisor: Exploring the role of
  large language models in fake news detection},''
  \emph{\BIBforeignlanguage{en}{Proceedings of the AAAI Conference on
  Artificial Intelligence}}, vol.~38, no. 2020, p. 22105–22113, Mar. 2024.
  [Online]. Available:
  \url{https://ojs.aaai.org/index.php/AAAI/article/view/30214}
\BIBentrySTDinterwordspacing

\bibitem{MaXea24_FakeNewsDetection}
\BIBentryALTinterwordspacing
X.~Ma, Y.~Zhang, K.~Ding, J.~Yang, J.~Wu, and H.~Fan, ``On fake news detection
  with llm enhanced semantics mining,'' in \emph{Proceedings of the 2024
  Conference on Empirical Methods in Natural Language Processing},
  Y.~Al-Onaizan, M.~Bansal, and Y.-N. Chen, Eds.\hskip 1em plus 0.5em minus
  0.4em\relax Miami, Florida, USA: Association for Computational Linguistics,
  Nov. 2024, p. 508–521. [Online]. Available:
  \url{https://aclanthology.org/2024.emnlp-main.31/}
\BIBentrySTDinterwordspacing

\bibitem{SpangherAea24_DoLLMsPlanLikeHumanWriters?}
\BIBentryALTinterwordspacing
A.~Spangher, N.~Peng, S.~Gehrmann, and M.~Dredze, ``Do llms plan like human
  writers? comparing journalist coverage of press releases with llms,'' in
  \emph{Proceedings of the 2024 Conference on Empirical Methods in Natural
  Language Processing}, Y.~Al-Onaizan, M.~Bansal, and Y.-N. Chen, Eds.\hskip
  1em plus 0.5em minus 0.4em\relax Miami, Florida, USA: Association for
  Computational Linguistics, Nov. 2024a, p. 21814–21828. [Online]. Available:
  \url{https://aclanthology.org/2024.emnlp-main.1216/}
\BIBentrySTDinterwordspacing

\bibitem{WuJea24_FakeNewsInSheepsClothing}
\BIBentryALTinterwordspacing
J.~Wu, J.~Guo, and B.~Hooi, ``Fake news in sheep’s clothing: Robust fake news
  detection against llm-empowered style attacks,'' in \emph{Proceedings of the
  30th ACM SIGKDD Conference on Knowledge Discovery and Data Mining}, ser. KDD
  ’24.\hskip 1em plus 0.5em minus 0.4em\relax New York, NY, USA: Association
  for Computing Machinery, Aug. 2024b, p. 3367–3378. [Online]. Available:
  \url{https://dl.acm.org/doi/10.1145/3637528.3671977}
\BIBentrySTDinterwordspacing

\bibitem{lai-etal-2023-chatgpt}
\BIBentryALTinterwordspacing
V.~D. Lai, N.~Ngo, A.~Pouran Ben~Veyseh, H.~Man, F.~Dernoncourt, T.~Bui, and
  T.~H. Nguyen, ``{C}hat{GPT} beyond {E}nglish: Towards a comprehensive
  evaluation of large language models in multilingual learning,'' in
  \emph{Findings of the Association for Computational Linguistics: EMNLP 2023},
  H.~Bouamor, J.~Pino, and K.~Bali, Eds.\hskip 1em plus 0.5em minus 0.4em\relax
  Singapore: Association for Computational Linguistics, Dec. 2023, pp.
  13\,171--13\,189. [Online]. Available:
  \url{https://aclanthology.org/2023.findings-emnlp.878}
\BIBentrySTDinterwordspacing

\bibitem{10.1145/3411763.3451760}
\BIBentryALTinterwordspacing
L.~Reynolds and K.~McDonell, ``Prompt programming for large language models:
  Beyond the few-shot paradigm,'' in \emph{Extended Abstracts of the 2021 CHI
  Conference on Human Factors in Computing Systems}, ser. CHI EA '21.\hskip 1em
  plus 0.5em minus 0.4em\relax New York, NY, USA: Association for Computing
  Machinery, 2021. [Online]. Available:
  \url{https://doi.org/10.1145/3411763.3451760}
\BIBentrySTDinterwordspacing

\bibitem{henkel2011reading}
L.~A. Henkel and M.~E. Mattson, ``Reading is believing: The truth effect and
  source credibility,'' \emph{Consciousness and cognition}, vol.~20, no.~4, pp.
  1705--1721, 2011.

\bibitem{WangWea19_ASurveyOfZero-ShotLearning}
\BIBentryALTinterwordspacing
W.~Wang, V.~W. Zheng, H.~Yu, and C.~Miao, ``A survey of zero-shot learning:
  Settings, methods, and applications,'' \emph{ACM Trans. Intell. Syst.
  Technol.}, vol.~10, no.~2, pp. 13:1--13:37, Jan. 2019. [Online]. Available:
  \url{https://dl.acm.org/doi/10.1145/3293318}
\BIBentrySTDinterwordspacing

\bibitem{BrownTea20_LLMsAreFewShotLearners}
\BIBentryALTinterwordspacing
T.~Brown, B.~Mann, N.~Ryder, M.~Subbiah, J.~D. Kaplan, P.~Dhariwal,
  A.~Neelakantan, P.~Shyam, G.~Sastry, A.~Askell, S.~Agarwal, A.~Herbert-Voss,
  G.~Krueger, T.~Henighan, R.~Child, A.~Ramesh, D.~Ziegler, J.~Wu, C.~Winter,
  C.~Hesse, M.~Chen, E.~Sigler, M.~Litwin, S.~Gray, B.~Chess, J.~Clark,
  C.~Berner, S.~McCandlish, A.~Radford, I.~Sutskever, and D.~Amodei, ``Language
  models are few-shot learners,'' in \emph{Advances in Neural Information
  Processing Systems}, vol.~33.\hskip 1em plus 0.5em minus 0.4em\relax Curran
  Associates, Inc., 2020, p. 1877–1901. [Online]. Available:
  \url{https://proceedings.neurips.cc/paper_files/paper/2020/hash/1457c0d6bfcb4967418bfb8ac142f64a-Abstract.html}
\BIBentrySTDinterwordspacing

\bibitem{wei2022chain}
J.~Wei, X.~Wang, D.~Schuurmans, M.~Bosma, F.~Xia, E.~Chi, Q.~V. Le, D.~Zhou
  \emph{et~al.}, ``Chain-of-thought prompting elicits reasoning in large
  language models,'' \emph{Advances in Neural Information Processing Systems},
  vol.~35, pp. 24\,824--24\,837, 2022.

\bibitem{KongAea24_BetterZeroShotReasoning}
\BIBentryALTinterwordspacing
A.~Kong, S.~Zhao, H.~Chen, Q.~Li, Y.~Qin, R.~Sun, X.~Zhou, E.~Wang, and
  X.~Dong, ``Better zero-shot reasoning with role-play prompting,'' in
  \emph{Proceedings of the 2024 Conference of the North American Chapter of the
  Association for Computational Linguistics: Human Language Technologies
  (Volume 1: Long Papers)}, K.~Duh, H.~Gomez, and S.~Bethard, Eds.\hskip 1em
  plus 0.5em minus 0.4em\relax Mexico City, Mexico: Association for
  Computational Linguistics, Jun. 2024, p. 4099–4113. [Online]. Available:
  \url{https://aclanthology.org/2024.naacl-long.228/}
\BIBentrySTDinterwordspacing

\bibitem{PelrineKea23_TowardsReliableMisinformationMitigation}
\BIBentryALTinterwordspacing
K.~Pelrine, A.~Imouza, C.~Thibault, M.~Reksoprodjo, C.~Gupta, J.~Christoph,
  J.-F. Godbout, and R.~Rabbany, ``Towards reliable misinformation mitigation:
  Generalization, uncertainty, and gpt-4,'' in \emph{Proceedings of the 2023
  Conference on Empirical Methods in Natural Language Processing}, H.~Bouamor,
  J.~Pino, and K.~Bali, Eds.\hskip 1em plus 0.5em minus 0.4em\relax Singapore:
  Association for Computational Linguistics, Dec. 2023, p. 6399–6429.
  [Online]. Available: \url{https://aclanthology.org/2023.emnlp-main.395/}
\BIBentrySTDinterwordspacing

\bibitem{json}
\BIBentryALTinterwordspacing
C.~Shorten, C.~Pierse, T.~B. Smith, E.~Cardenas, A.~Sharma, J.~Trengrove, and
  B.~van Luijt, ``Structuredrag: Json response formatting with large language
  models,'' 2024. [Online]. Available: \url{https://arxiv.org/abs/2408.11061}
\BIBentrySTDinterwordspacing

\bibitem{schulhoff2024prompt}
\BIBentryALTinterwordspacing
S.~Schulhoff, M.~Ilie, N.~Balepur, K.~Kahadze, A.~Liu, C.~Si, Y.~Li, A.~Gupta,
  H.~Han, S.~Schulhoff, P.~S. Dulepet, S.~Vidyadhara, D.~Ki, S.~Agrawal,
  C.~Pham, G.~Kroiz, F.~Li, H.~Tao, A.~Srivastava, H.~D. Costa, S.~Gupta, M.~L.
  Rogers, I.~Goncearenco, G.~Sarli, I.~Galynker, D.~Peskoff, M.~Carpuat,
  J.~White, S.~Anadkat, A.~Hoyle, and P.~Resnik, ``The prompt report: A
  systematic survey of prompting techniques,'' 2024. [Online]. Available:
  \url{https://arxiv.org/abs/2406.06608}
\BIBentrySTDinterwordspacing

\bibitem{zhou2023leasttomostpromptingenablescomplex}
\BIBentryALTinterwordspacing
D.~Zhou, N.~Schärli, L.~Hou, J.~Wei, N.~Scales, X.~Wang, D.~Schuurmans,
  C.~Cui, O.~Bousquet, Q.~Le, and E.~Chi, ``Least-to-most prompting enables
  complex reasoning in large language models,'' 2023. [Online]. Available:
  \url{https://arxiv.org/abs/2205.10625}
\BIBentrySTDinterwordspacing

\bibitem{SuzgunMea23_ChallengingBIGBenchTasks}
\BIBentryALTinterwordspacing
M.~Suzgun, N.~Scales, N.~Schärli, S.~Gehrmann, Y.~Tay, H.~W. Chung,
  A.~Chowdhery, Q.~Le, E.~Chi, D.~Zhou, and J.~Wei, ``Challenging big-bench
  tasks and whether chain-of-thought can solve them,'' in \emph{Findings of the
  Association for Computational Linguistics: ACL 2023}, A.~Rogers,
  J.~Boyd-Graber, and N.~Okazaki, Eds.\hskip 1em plus 0.5em minus 0.4em\relax
  Toronto, Canada: Association for Computational Linguistics, Jul. 2023, p.
  13003–13051. [Online]. Available:
  \url{https://aclanthology.org/2023.findings-acl.824/}
\BIBentrySTDinterwordspacing

\bibitem{kojima2022large}
T.~Kojima, S.~S. Gu, M.~Reid, Y.~Matsuo, and Y.~Iwasawa, ``Large language
  models are zero-shot reasoners,'' \emph{Advances in neural information
  processing systems}, vol.~35, pp. 22\,199--22\,213, 2022.

\bibitem{ji-etal-2023-towards}
\BIBentryALTinterwordspacing
Z.~Ji, T.~Yu, Y.~Xu, N.~Lee, E.~Ishii, and P.~Fung, ``Towards mitigating {LLM}
  hallucination via self reflection,'' in \emph{Findings of the Association for
  Computational Linguistics: EMNLP 2023}, H.~Bouamor, J.~Pino, and K.~Bali,
  Eds.\hskip 1em plus 0.5em minus 0.4em\relax Singapore: Association for
  Computational Linguistics, Dec. 2023, pp. 1827--1843. [Online]. Available:
  \url{https://aclanthology.org/2023.findings-emnlp.123}
\BIBentrySTDinterwordspacing

\bibitem{corso2024conspiracy}
F.~Corso, F.~Pierri, and G.~D.~F. Morales, ``Conspiracy theories and where to
  find them on tiktok,'' \emph{arXiv preprint arXiv:2407.12545}, 2024.

\end{thebibliography}


\begin{thebibliography}{1}
\providecommand{\url}[1]{#1}
\csname url@samestyle\endcsname
\providecommand{\newblock}{\relax}
\providecommand{\bibinfo}[2]{#2}
\providecommand{\BIBentrySTDinterwordspacing}{\spaceskip=0pt\relax}
\providecommand{\BIBentryALTinterwordstretchfactor}{4}
\providecommand{\BIBentryALTinterwordspacing}{\spaceskip=\fontdimen2\font plus
\BIBentryALTinterwordstretchfactor\fontdimen3\font minus
  \fontdimen4\font\relax}
\providecommand{\BIBforeignlanguage}[2]{{%
\expandafter\ifx\csname l@#1\endcsname\relax
\typeout{** WARNING: IEEEtran.bst: No hyphenation pattern has been}%
\typeout{** loaded for the language `#1'. Using the pattern for}%
\typeout{** the default language instead.}%
\else
\language=\csname l@#1\endcsname
\fi
#2}}
\providecommand{\BIBdecl}{\relax}
\BIBdecl

\bibitem{Quelle_2024}
\BIBentryALTinterwordspacing
D.~Quelle and A.~Bovet, ``The perils and promises of fact-checking with large
  language models,'' \emph{Frontiers in Artificial Intelligence}, vol.~7, Feb.
  2024. [Online]. Available: \url{http://dx.doi.org/10.3389/frai.2024.1341697}
\BIBentrySTDinterwordspacing

\end{thebibliography}

\end{document}